\documentclass[]{aa}
\usepackage{graphicx}
\usepackage[varg]{txfonts}
\usepackage{url}
\newcommand{\kms}{km\,s$^{-1}$}

\usepackage{multirow}
\usepackage{color}
\usepackage{pdflscape}
\begin{document} 
\title{Molecular remnant of Nova 1670 (CK Vulpeculae)}
\subtitle{II. A three-dimensional view on the gas distribution and velocity field}
   \author{T. Kami\'nski \inst{1,2}
           \and
          W. Steffen\inst{3}
           \and
          V. Bujarrabal\inst{4}
          \and
          R. Tylenda\inst{1}
           \and
          K. M.\ Menten\inst{5}
          \and
          M. Hajduk\inst{6}
          }

\institute{\centering
Nicolaus Copernicus Astronomical Center, Polish Academy of Sciences, ul. Rabia{\'n}ska 8, 87-100 Toru{\'n}, Poland, \email{tomkam@ncac.torun.pl}\label{inst1}
\and
Center for Astrophysics $|$ Harvard \& Smithsonian , 60 Garden Street, Cambridge, MA 02138, USA \label{inst2}
\and
Instituto de Astronom\'ia, OAN, UNAM, Ensenada, Mexico\label{inst3}
\and
Observatorio Astronomico Nacional (OAN, IGN), Campus Universitario, Ctra. NII km 33\,600, Apartado 112, E-28803 Alcala de Henares, Spain\label{inst4}
\and
Max Planck Institut f\"ur Radioastronomie, Auf dem H\"ugel 69, D-53121 Bonn, Germany\label{inst5}
\and
University of Warmia and Mazury, Faculty of Geoengineering, ul. Prawochenskiego 15, 10-720 Olsztyn, Poland\label{inst6}
}
\authorrunning{T. Kami\'nski et al.}
\abstract{CK\,Vul is the remnant of an energetic eruption known as Nova 1670 that is thought to be caused by a stellar merger. The remnant is composed of (1) a large hourglass nebula of recombining gas (of 71\arcsec size), very similar to some classical planetary and pre-planetary nebulae (PPNe), and (2) of a much smaller and cooler inner remnant prominent in millimeter-wave emission from molecules. We investigate the three-dimensional spatio-kinematic structure of both components. The analysis of the hourglass structure yields a revised distance to the object of $>$2.6\,kpc, at least 3.7 times greater than so far assumed. At this distance, the stellar remnant has a bolometric luminosity >12\,L$_{\sun}$ and is surrounded by molecular material of total mass $>$0.8\,M$_{\sun}$ (the latter value has a large systematic uncertainty). We also analyzed the architecture of the inner molecular nebula using ALMA observations of rotational emission lines obtained at subarcsecond resolution. We find that the distribution of neutral and ionized gas in the lobes can be reproduced by several nested and incomplete shells or jets with different velocity fields and varying orientations. The analysis indicates that the molecular remnant was created in several ejection episodes, possibly involving an interacting binary system. We calculated the linear momentum ($\approx$10$^{40}$\,g\,cm\,s$^{-1}$) and kinetic energy ($\approx$10$^{47}$\,erg) of the CK\,Vul outflows and find them within the limits typical for classical PPNe. Given the similarities of the CK\,Vul outflows to PPNe, we suggest there may CK\,Vul analogs among wrongly classified PPNe with low intrinsic luminosities, especially among PPNe with post-red-giant-branch central stars.}

\maketitle

\section{Introduction}\label{intro}
Nebulae of post asymptotic-giant-branch (post-AGB) stars have been extensively imaged at optical, infrared, and millimeter wavelengths. A large fraction of them has a  bipolar shape. Bipolar planetary nebulae (PNe) and pre-planetary nebulae (PPNe) are shaped by mechanisms that are not entirely understood \citep{BalickFrank}. Magnetic fields and stellar rotation may be involved in producing some asymmetries of the nebulae but are not considered to be the dominant actors. The still elusive sculpting mechanisms most likely involve binary interactions \citep[e.g.,][]{BujarrabalDiscussion,SokerRappaport}, among them common-envelope ejections and mergers \citep{Paczynski76,JonesBoffin,SokerKashi2012,submmRN}. Bipolar nebulae display a wide range of morphologies and kinematical properties. These include structures formed in multiple ejections with different orientations \citep[e.g.,][]{SahaiTrauger}, atomic and molecular outflows expanding with high speeds (molecular with roughly 100--300\,\kms\ and atomic up to about 2000\,\kms) \citep[e.g.][]{fastAtomic} and following the Hubble law, that is, with the speed increasing linearly with distance from the origin. In addition, rotating as well as expanding disks and tori in the waist of the bipolar structure have been observed \citep{Bujarrabal2016,Bujarrabal2018,SanchezContreras2000}. The momenta of the observed outflows nearly always exceed by orders of magnitude values expected from radiation pressure of the luminous central star. Thus, the origin of the bipolar PPNe must be related to dynamic phenomena \citep[e.g.,][]{Alcolea2001}, possibly involving accretion episodes in binary or multiple systems \citep{SanchezContreras2004}. Most well-studied bipolar PPNe and PNe have kinematic ages of a few hundred years ($\lesssim$1500\,yr) and it is suspected that they had been created in a relatively short phase, lasting tens to a few hundred years \citep{Bujarrabal2001}. Many of the nebulae encompass bullets or jets \citep[see e.g.,][]{Huang2016} (although the definition of a jet varies between authors). Identifying the shaping mechanism of those post-AGB stars is important for our understanding of the  stellar evolution beyond the main sequence in low-mass binary and multiple systems. 

The object CK\,Vul, when observed at optical wavelengths, appears very similar to many bipolar PPNe: it has a large (71\arcsec) hourglass nebula of recombining gas, multiple ``bullets'' ejected at different orientations \citep{hajduk2007}, a dust-rich waist \citep{nature}, and fast atomic and molecular outflows \citep{nature,hajduk2013} which follow the Hubble law. What clearly distinguishes CK\,Vul from post-AGB systems is that its central object is much less luminous ($\gtrsim$0.9\,L$_{\sun}$ in \citealp{nature}; revised here to 16--60\,L$_{\sun}$) than post-AGB stars (typically of 10$^4$\,L$_{\sun}$ for PPNe). More importantly, in contrast to  PPNe and PNe, CK\,Vul has been observed directly in outburst. The ``new star'' was discovered in June 1670 and its changing visual brightness was extensively documented until its final dimming sometime after May 1672. Owing to the work of  17th-century observers --- including P\'ere Dom Anthelme, Johannes Hevelius, and Giovanni Domenico Cassini --- \citet{shara85} were able to reconstruct the light curve of Nova 1670. There had been at least three major outbursts over the $\approx$3\,yr that the {\it nova} was visible to the naked eye. Such a light curve is not typical for classical novae \citep{Rosenbush}, but is very typical of a recently-recognized group of eruptive objects known as (luminous) red novae or (intermediate-luminosity) red transients. A reddish color of the 1670 transient was noted by J. Hevelius \citep{shara85}. Red color is a tell-tale characteristic of red novae in late stages of their eruptions, due to dramatic cooling and increasing obscuration by newly-formed dust. The ancient eruption was therefore postulated to be of the red-nova kind \citep{Kato,TylendaBLG}. That hypothesis was corroborated by further analysis of the remnant \citep{nature,NatAstr,kami-singledish}. (Some authors have presented a different view on the nature of the object, see e.g., \cite{shara85}, \cite{MB}, \cite{evans2016}, and \cite{Eyres}, but most of them are invalided by more recent observations; see e.g., \citealp{nature} and \citealp{NatAstr}). As argued by \citet{TylendaSoker2006} and directly observed for the case of V1309\,Sco \citep{Tylenda2011}, red novae erupt as the results of stellar mergers, events in which the coalescence is preceded by a quick phase of the common-envelope evolution \citep{LivioSoker,IvanovaReview,MacLeod2017}. Interestingly, some of the red transients, including extragalactic ones, have been linked to PPNe \citep{Prieto,SokerKashi2012}. The case of Nova 1670 and its remnant, CK\,Vul, makes this link even stronger.

\begin{figure}\sidecaption
  \includegraphics[trim=20 90 20 65, clip, width=\columnwidth]{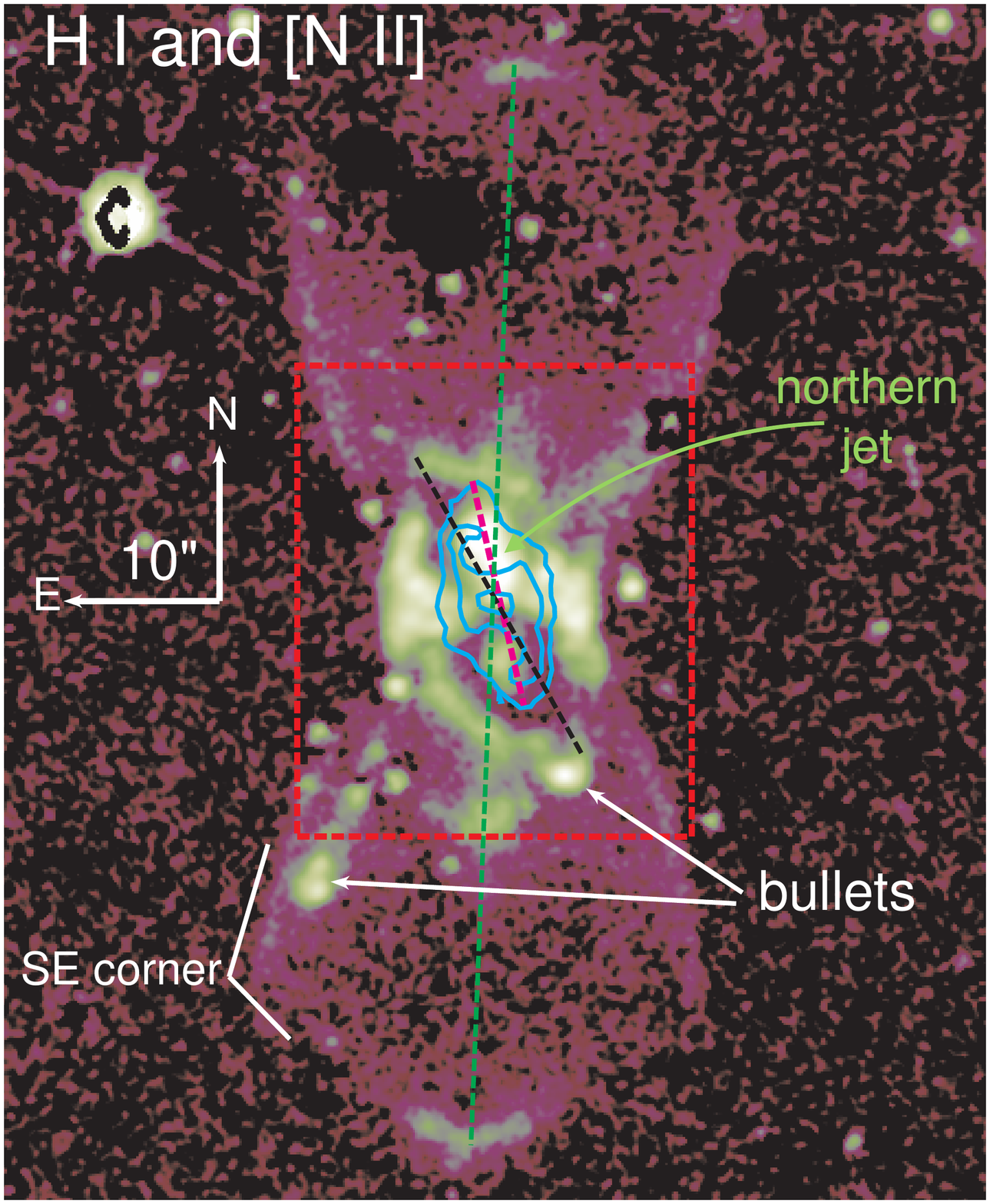}
\caption{Large-scale hourglass nebula of CK\,Vul. The background optical image show the nebular emission in lines of H$\alpha$ and [\ion{N}{ii}]. Although most stars are removed from the image, some considerable residuals can be seen (most of the point-like sources) \citep{hajduk2007}. The red dashed rectangle of 26\arcsec$\times$31\arcsec\ corresponds to the size of the maps in Fig.\,\ref{fig-co-opt}. The cyan contours show emission of CO 1--0. The dashed lines mark the symmetry axes of: the large hourglass nebula (green), the CO region (black), and the optical shell immediately surrounding the molecular nebula (magenta).} \label{fig-co-opt-largescale}
\end{figure}

The nebular remnant of Nova 1670, was recovered only quite recently by \citet{shara82}. Subsequent observations showed the presence of the large hourglass nebula within which several clumps of atomic emission were found by \cite{hajduk2007}, including features which we call here the northern jet and bullets (Fig.\,\ref{fig-co-opt-largescale}). The photosphere of the stellar remnant has not been observed, neither directly nor in scattered light, even though very sensitive observations of the center of the nebula have been performed \citep{shara85,Naylor,hajduk2007,hajduk2013,xshooter}. The star is obscured by dust at visual and infrared wavelengths and only at long radio wavelengths a compact source was found, presumably manifesting ionized material located in the immediate vicinity of the photosphere \citep{hajduk2007}. 

Quite surprising were observations of CK\,Vul at millimeter (mm) and submillimeter (submm) wavelengths that have been conducted over the last few years. They revealed a very rich molecular component of the remnant. It is composed of simple astrophysically-common di- and triatomics (such as CO, HCN, SiO), but also contains polyatomic molecules, as complex as CH$_3$OH and CH$_3$NH$_2$, the latter being not too common in the interstellar medium (ISM) and absent in the envelopes of evolved stars and CH$_3$OH only observed toward very few such objects \citep{nature,kami-singledish}. The molecular gas is enhanced in helium, products of the CNO cycles, and partial helium burning, including rare CNO isotopes and $^{26}$Al. Their presence is consistent with a red giant branch (RGB) star torn apart in the merger of 1670 \citep{nature,kami-singledish,NatAstr}. The molecular gas forms a bipolar structure that is approximately seven times smaller than the atomic hourglass nebula. 
Partial analysis of the molecular remnant has been presented in \cite{nature}, \cite{NatAstr}, and \cite{Eyres}. While proper motions of the atomic nebula seem to indicate that its kinematic age is consistent with ejection in the eruption of 1670--72 \citep{hajduk2007}, the age of the molecular bipolar structure is unknown. It has been suggested that it could have been created long after 1670 \citep{Kami2020,Eyres} but solid evidence for this is missing. 

In \citet[][hereafter Paper\,I]{Kami2020}, we presented an imaging line survey conducted with the Atacama Large Millimeter/submillimeter Array (ALMA) and Submillimeter Array (SMA) interferometers. The survey allowed us to explore the spatial complexity of the molecular remnant that was traced in 180 spectral lines of 22 molecular species. The bipolar lobes and the central region are well resolved by the interferometers, thereby allowing first detailed studies of the remnant's components. Non-local thermodynamic equilibrium (non-LTE) analysis of the extended lobes yield a gas temperature of $\sim$14\,K and densities of 10$^5$\,cm$^{-3}$. The central region, in the waist of the molecular and atomic nebulae, is slightly denser (10$^6$\,cm$^{-3}$) and warmer, with a bulk gas temperature of $\sim$17\,K, and characterized by a temperature gradient. The central region appears to be richer in oxygen-bearing molecules than the lobes. The presence of the more complex species, which are chiefly observed in the lobes, is difficult to understand from an astrochemical point of view: given the environment, it seems natural that  shock waves were driving their creation, either via the destruction of dust grains and their icy mantles (if any existed) or in endothermic reactions in the hot post-shock regions; the latter processes have to date been little explored by astrochemists. Quadrupole transitions of H$_2$ arising from a clump within the molecular nebula (Paper\,I) and the excitation of atomic lines \citep{xshooter} are also indicative of fast (50--90\,\kms) active shocks. The gas in the lobes displays very complex kinematics but, in general, can be understood if the lobes are part of an hourglass structure oriented at a very small inclination angle to the sky plane. (For inclination, we adopt the convention where it is the angle between the long axis of the hourglass and the sky plane). 

In this paper, we take a closer look at the three-dimensional spatio-kinematic structure of the atomic and molecular remnant of Nova 1670. Observational data at millimeter wavelengths have been described and presented in Paper\,I. We start the current study with a revision of the basic astrophysical parameters of the remnant, including the distance and the luminosity, based on optical observations of the large hourglass nebula. The revision is presented in Sect.\,\ref{sec-hourglass}. In Sect.\,\ref{sec-modeling}, we turn mainly to the inner remnant and introduce our modeling methods aimed at reproducing its architecture in 3D. We present results of the modeling for different components of the remnant, that is, for bulk neutral gas (Sects.\,\ref{sec-outer} and \ref{sec-3D-neutral}), ionized molecular gas (Sect.\,\ref{sec-model-ions}), and for an extended CO-bearing region (Sect.\,\ref{sec-bubbles}). In Sect.\,\ref{sec-ekin}, we constrain the linear momentum and kinetic energy of the molecular outflow. Finally, in Sect.\,\ref{sec-discussion}, we discuss the results in search for clues on the dynamical origin of the molecular outflow in CK\,Vul and the object's relation to the known PNe and PPNe. 

\section{Revised physical parameters of the remnant}\label{sec-hourglass}
Distance and luminosity of the remnant are essential for revealing the nature of CK\,Vul. The distance of 0.7\,kpc has often been assumed for CK\,Vul after \citet{hajduk2013}. This distance was derived from an analysis of the large hourglass nebula that can be safely assumed to originate in the eruption of 1670. Based on images and long-slit spectroscopy of the nebula, Hajduk et al. performed a 3D analysis which lead them to the inclination of the structure. Their figures present a model for an inclination of 25\degr. Knowing the inclination and radial velocities, and assuming a simple velocity field such as the Hubble flow, one can derive the tangential and the total deprojected velocities of the gas. Then, the angular size of the structure gives simply the distance to the object. In calculating the distance, Hajduk et al., however, used an inclination of 65\degr\ (sic), inconsistent with the shown model (with 25\degr), and found a distance of 700\,pc. We correct this error and propose a revised model of the hourglass nebula that implies a much greater distance.

\begin{figure} \centering 
  \includegraphics[width=0.28\columnwidth, trim=135 0 135 0, clip]{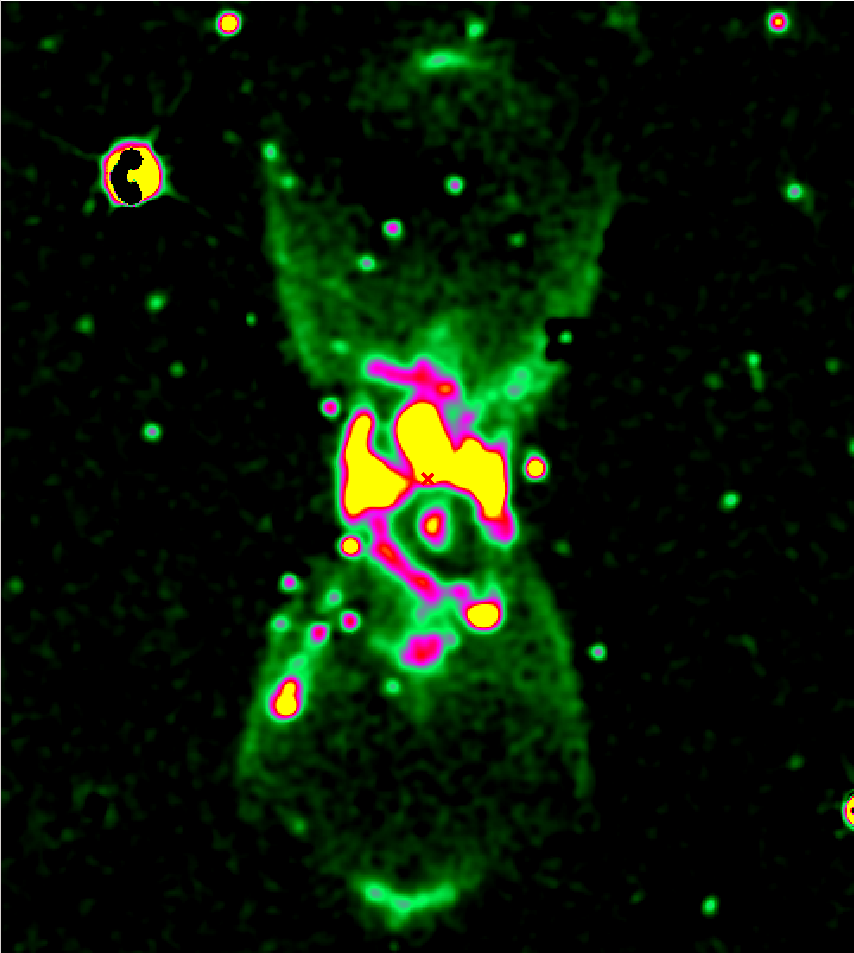}
  \includegraphics[width=0.31\columnwidth, trim=95 20 95 20, clip]{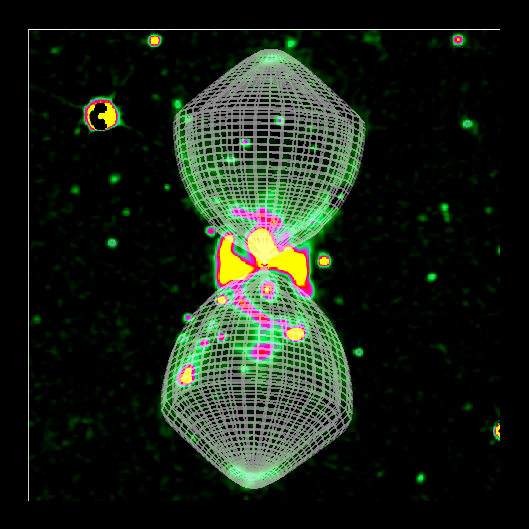}
  \includegraphics[width=0.302\columnwidth, trim=70 15 70 15, clip]{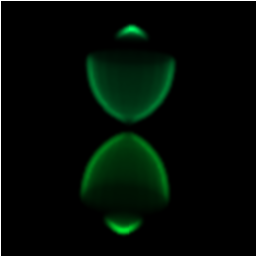}
\caption{Three-dimensional model of the atomic hourglass nebula of CK\,Vul. Left: The observed image of the nebula (same as in Fig.\,\ref{fig-co-opt-largescale}). Bright nebula regions shown in yellow and magenta are  not considered to be parts of the hourglass. Center: The bipolar-shell model rendered as a mesh is overplotted on the observed image. Right: Simulated image of the nebula for our line of sight. Polar regions have enhanced emissivity while emission adjacent to it was suppressed.} \label{fig-model-hourglass}
\end{figure}

We used the SHAPE package \citep[][version 5.1]{SL,steffen} to reconstruct the overall spatio-kinematic structure of the large hourglass nebula and compared it to images and long-slit spectra presented in \citet{hajduk2013}. We assumed that the shape and size of the nebula are the same along the line of sight as in the sky plane. Each lobe was modeled as a single shell-like structure. In particular, pointy (nearly triangular) cusps of emission seen at the northern and southern tips of each lobe (Fig.\,\ref{fig-co-opt-largescale}) were assumed to be part of the lobes. In the modeling and comparison to the observations, we ignored the inner nebula regions up to radii of about 7\arcsec, where the hourglass structure overlaps with multiple bright features of unknown kinematical ages (cf. Fig.\,\ref{fig-model-hourglass}). We assumed a simple Hubble-flow expansion of the shells. The relative brightness distribution of the hourglass was only roughly reproduced (as a function of the polar angle) since it has little meaning for the distance determination. The appearance of the nebula was simulated assuming optically thin emission.

The characteristic presence of the two clear bright cusps at the tips of the hourglass nebula and the simultaneous presence of sharp angles at the NE, NS, SE, and SW ``corners'' of the hourglass (cf. Fig.\,\ref{fig-co-opt-largescale}) imply an upper limit of $\approx$35\degr\ on the inclination. At larger inclinations, the tips of the lobes appear internal to the lobes and instead of the corners, the rims of the hourglass appear rounded for a wide range of intrinsic shapes. In particular, an inclination of 65\degr\ used by \citet{hajduk2013} to yield the distance can be practically ruled out if the cusps are an integral part of the lobes. The nebula morphology does not provide us with a lower limit on the inclination. In Fig.\,\ref{fig-model-hourglass}, we present a sample model where the northern lobe has an inclination of 10\degr\ and the southern one is seen at 20\degr\ inclination. Such a model reproduces very well the spectra obtained with a slit located at the high-proper motion star west of CK\,Vul and crossing eastern walls of both lobes \citep[cf. Figs.\,3 and 4 in][]{hajduk2013}. We find some subtle asymmetries between the southern and northern lobes (e.g., in size, position and inclinations angles) but these will be explored in more length and compared to more sensitive observations in a forthcoming paper (Hajduk et al., in preparation).

In spectra of the bright [\ion{N}{II}] $\lambda$6583 line, the apexes of the hourglass have an average radial velocity of $\pm$820\,\kms\ (blueshifted in the south and redshifted in the north). With the radial extent of each lobe of 36\arcsec\ and the kinematical age of 340\,yr, we calculated the distance and other physical parameters of CK\,Vul's remnant for different inclination angles. Results are shown in Table\,\ref{tab-distance}. Our upper limit of $\approx$35\degr\ on the inclination implies that the object is at distances larger than $\approx$2.6\,kpc. Our show-case model with $i$=15\degr$\pm$5\degr\ from Fig.\,\ref{fig-model-hourglass} puts CK\,Vul at a distance of 5.7\,kpc and indicates the ejecta velocity of 3200\,\kms. That is very fast but not unreasonable. We consider this solution as a sensible upper limit on the distance. With the object's Galactic longitude of $l$=63\fdg3,  at much lower inclination angles (for instance of 5\degr) the distances would be enormous (17\,kpc) putting CK\,Vul at the outskirts of the Galaxy and would imply gas motions comparable to those of supernovae. Such solutions are therefore highly unlikely. 

An independent constraint on the distance to CK\,Vul can be derived from the distance to a field star named variable 2 \citep{hajduk2013} that is located behind the southern molecular lobe. Its Gaia Data Release 2 parallax \citep{gaia} implies a distance of 4.9\,kpc or a 1$\sigma$-like confidence interval 3.0--8.1\,kpc \citep[calculated after][]{Bailer} and implying 10\degr<$i$<30\degr. 
We can thus quite confidently claim that the distance to CK\,Vul is $>$2.6\,kpc and we adopt 5.7\,kpc as a plausible upper limit. A distance of 2.6 (5.7) kpc would place CK\,Vul 45 (98) pc above the Galactic plane, well withing the thin disk.

\begin{table}
    \caption{Physical parameters of the remnant at different inclinations.}\label{tab-distance}
    \centering\footnotesize
    \begin{tabular}{ccccc}
\hline 
$i$ &  $V_{\rm exp}$ & $d$   & $L_{\rm bol,now}$ & $L_{\rm bol,max}$\\  
(\degr)&(\kms)       & (kpc) & ($L_{\sun}$)  &  ($L_{\sun}$) \\ 
\hline  
\hline
 5 & 9413 & 17.0 & 532 & 2.7e8\\
15 & 3170 & 5.7 &  60 & 3.0e7\\
25 & 1941 & 3.5 &  23 & 1.1e7\\
35 & 1420 & 2.6 &  12 & 6.2e6\\
65? & 900? & 0.7 & 0.9 & 4.5e5\\
\hline 
    \end{tabular}
    \tablefoot{$V_{\rm exp}$ is the total (deprojected) expansion speed of the nebula. $L_{\rm bol,now}$ represents the contemporary bolometric luminosity of the object and $L_{\rm bol,max}$ is the bolometric luminosity of Nova 1670 during its maximum visual brightness in 1671. The last row corresponds to values given in \citet{hajduk2013}. Some values are given with a question mark to indicate they are not consistent with the provided distance.}
\end{table}

The bolometric luminosity of CK\,Vul's remnant was derived on the basis of a spectral energy distribution (SED) of the object \citep{nature}. The SED was reconstructed from broad-band photometric measurements ranging from infrared to radio wavelengths. That luminosity may be somewhat underestimated because it does not take into account radiation escaping the system through polar regions. At the assumed distance of 0.7\,kpc, a luminosity of 0.9\,L$_{\sun}$ was derived. The revised distance makes the remnant of CK\,Vul much more luminous. From the above analysis, the bolometric luminosity must be >12\,L$_{\sun}$. If CK\,Vul is located in the Galactic disk, its luminosity is certainly lower than 10$^4$\,L$_{\sun}$. The revised luminosity of CK\,Vul matches well those of RGB or post-RGB stars, consistent with the RGB merger hypothesis of \cite{NatAstr}.

The revised distance also implies a high luminosity of the outburst of Nova 1670. We calculated its maximal bolometric luminosity for the range of distances and present them in the last column of Table\,\ref{tab-distance}. In the calculations, we adopted: the minimal visual magnitude of 2.6 during the 1671 peak \citep{shara85}; interstellar visual extinction of $A_V$=2.79\,mag \citep[an upper limit from][]{xshooter}; and a bolometric correction of $0.0$\,mag consistent with a supergiant at a temperature of $\sim$7500\,K \citep[i.e., comparable to that of V838\,Mon, cf.][]{TylendaV838}. Although the assumed interstellar extinction and the bolometric correction are uncertain, the peak luminosity of Nova\,1670 was very likely on the order of $10^7$\,L$_{\sun}$, perhaps exceeding that of the brightest Galactic red-nova outburst of V838\,Mon \citep{sparks} but comparable to peak luminosities of some extragalactic merger-burst candidates \citep[cf.][]{Pastorello}.

\section{Three-dimensional molecular remnant}\label{sec-modeling}
With the revised basic parameters of the remnant, we come back to the molecular component of the remnant. We again used the SHAPE package (version X.$\alpha$.619 from 2018) to reconstruct the overall spatio-kinematic structure in 3D.
Our simulations were performed within the optically-thin limit and we assumed a homologous expansion, as expected from an eruptive event with ballistic ejections. Cylindrical or point-symmetry was assumed unless otherwise stated. 

SHAPE models were iteratively constructed by comparing the simulated spatio-kinematic structure to products of the observational data, including  total-intensity maps, channel-velocity maps, and position-velocity (PV) diagrams generated for several virtual slits and multiple representative molecules. The observables were extracted from average data-cubes introduced in Paper\,I. They typically represent a noise-weighted mean of multiple transitions and isotopologues of a given species (except for CO; see below). Most of the observational material discussed in the current study is based on the more sensitive ALMA observations.
In SHAPE, we worked with an arbitrary flux scale focusing on reconstructing the relative intensities seen in the interferometric maps.  

The molecular emission is generally optically thin in the bipolar lobes on which we focus here, while we 
omit the ``central region''. Defined in Paper\,I, the central region is within $\approx$2\arcsec\ of the remnant center. It is characterized by a considerable range of temperatures, optically thick emission of the most abundant species, and has an irregular velocity field traced in SO$_2$ lines (see Paper\,I). To address all this, would require adopting a different modeling approach than that applied here. 

The morphology of the lobes is shown for a number of molecules in Fig.\,2 of Paper\,I. The lobes have a very pronounced point symmetry, but small asymmetries between the northern and southern lobes can be noticed. In the majority of the observed molecules, the nebula exhibits an S-shape morphology, except for molecular ions, which have an \reflectbox{S}-shape (inverse-S) appearance in total intensity maps (see Fig.\,\ref{fig-slits-ions}).  The emission distributions of neutral and ionic species complement each other to form an 8-shaped structure. 

\subsection{Step 1: General shape}\label{sec-outer}
\begin{figure}
  \includegraphics[width=.99\columnwidth]{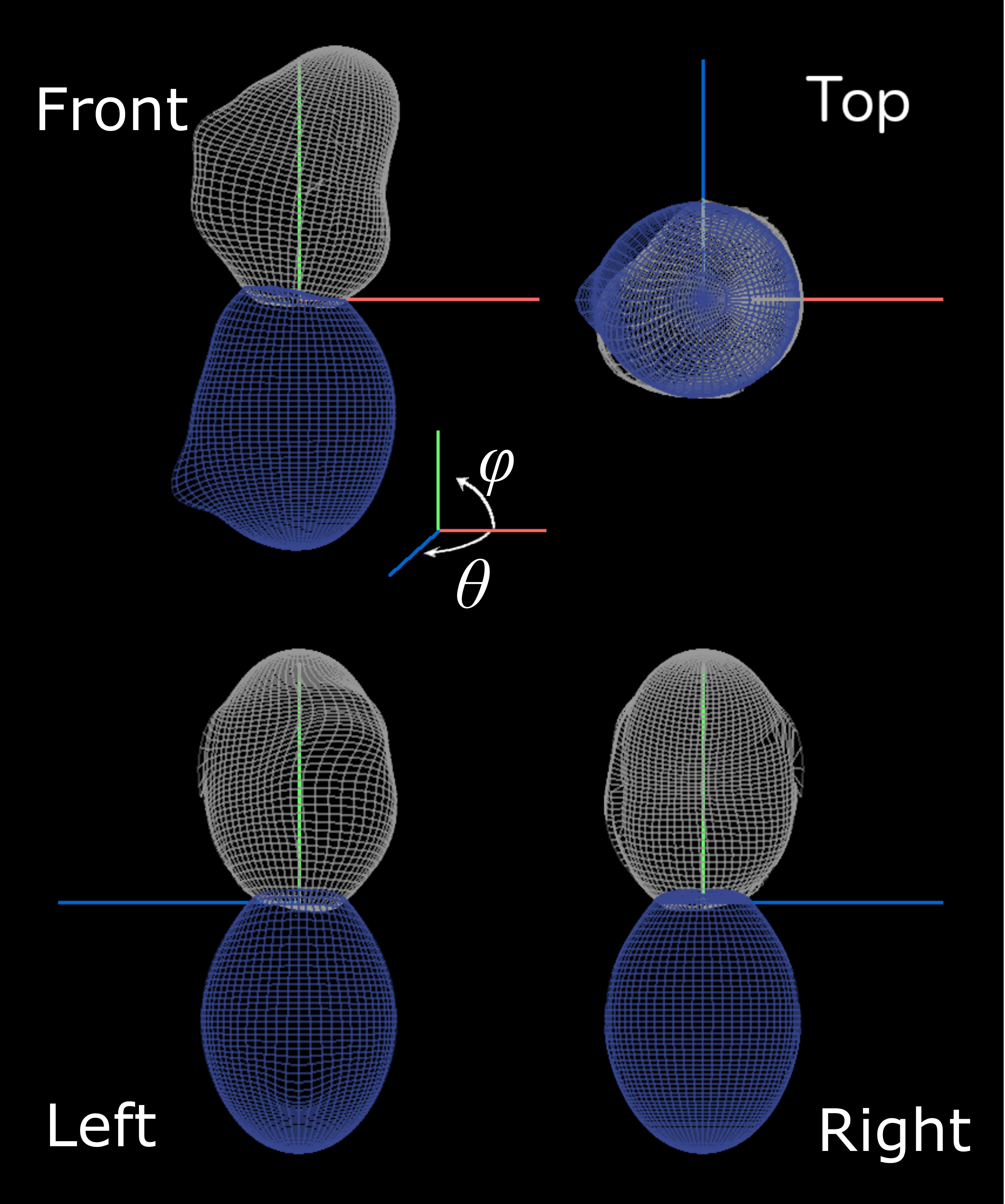}
\caption{General shape of the molecular lobes in our 3D reconstruction. The outer shells are represented by meshes viewed from four angles. The front view is that for a terrestrial observer but, for convenience, rotated by a position angle of --17\degr. Northern lobe is shown with a white mesh, southern with a blue one. Cartesian $xyz$ axes are shown for reference in red, green, and blue, respectively. The blue axis corresponds to the line of sight. The convention of counting the azimuthal ($\theta$) and polar ($\phi$) angles is shown in the inset.  Each lobe is $\sim$8\arcsec\ long.}\label{fig-shape1}
\end{figure}

First, we focused on reproducing the shape and kinematics of the outermost spatial structures (those farthest from the expansion center) that are visible in neutral-gas tracers observed at a very high signal-to-noise ratio (S/N), such as in HCN and SiO. Each lobe was represented by an ellipsoidal structure whose shape was next manually modified by stretching and squeezing, and by adding several distortions called ``bumps'' in SHAPE. The sizes of the structures along the sight-line were assumed to be similar to their widths measured directly in the maps. The overall simulated shape of the outer remnant is shown in Fig.\,\ref{fig-shape1}. 

The velocity field was implemented as a stitched function of linear radial expansion laws, in the form $v(r)=kr$ and with different values of $k$. The reference frame of the velocity field was oriented in the same manner as the spatial structure and thus defined by the same inclination and position angles. For the bulk of molecular gas in these outer shells, we were able to satisfactorily reproduce the observations with $k$=37\,\kms\,arcsec$^{-1}$ near the central region (i.e., roughly within a radius, $r$, of 1\arcsec) and $k$=58--68\,\kms\,arcsec$^{-1}$ outwards. In our preferred implementation, the lobes are only slightly inclined to the sky plane: the northern lobe is at an inclination $i$=10\degr\ and the southern one is in the sky plane ($i$=0). The values of $k$ and $i$ are degenerate and cannot be determined independently. Thus, our model does not generally offer a unique representation of the available data. However, high inclinations, say >30\degr, are highly unlikely as they would impose spatio-kinematic features inconsistent with observations. For example, it is certain that the inclination is smaller than the opening angles of the lobes because we observe both redshifted and blue-shifted emission from each lobe (cf. Paper\,I). A small asymmetry in the velocity ranges with respect to the systemic LSR velocity of $-10$\,\kms\ is observed and indicates that the gas is slightly more redshifted in the northern lobe. In particular, spectra extracted at the tips of the molecular nebula are redshifted in the north and blueshifted in the south. We note that this orientation is opposite to that seen in the cusps of the hourglass structure \citep{hajduk2013}.

Within this model, the outermost emission is located at 7\farcs2 where gas moves with a deprojected speed of 470\,\kms. Adopting a distance of $3.5^{+2.1}_{-0.9}$\,kpc (Sect.\,\ref{sec-hourglass}), ballistic motion with such a speed would take $254^{+153}_{-65}$\,yr to reach the outer parts of the molecular remnant. This, within the uncertainties, is consistent with an origin of the lobes close to the 1670--72 eruption. Slightly higher inclinations require larger shells expanding with considerably lower speeds. For example, at an inclination near 30\degr, the shells must be $\sim$15\% longer and 50\% slower resulting in over twice higher kinematical age, that is 585\,yr at 3.5\,kpc. The upper limit on the kinematical age, limits the range of inclinations to angles lower than about 15\degr. (This should not be confused with the inclination of the large hourglass structure.)


Assuming that the main axes of the molecular lobes are in the sky plane, as nearly is the case for our 3D model, the kinematical age of the outer shells may also be determined using solely the maximal radial motions of the gas within this structure, that is 134\,\kms. We assume that the extension of the lobes along the line of sight is the same as the size in the sky plane, that is with a radius of $\approx$2\farcs5. At an age of 340\,yr, this indicates a distance of 3.8\,kpc, which is consistent with the revised kinematical distance to the large hourglass optical nebula (Sect.\,\ref{sec-hourglass}).

\subsection{Step 2: Distribution of neutral species}\label{sec-3D-neutral}

\begin{figure}\centering
  \includegraphics[width=.49\columnwidth]{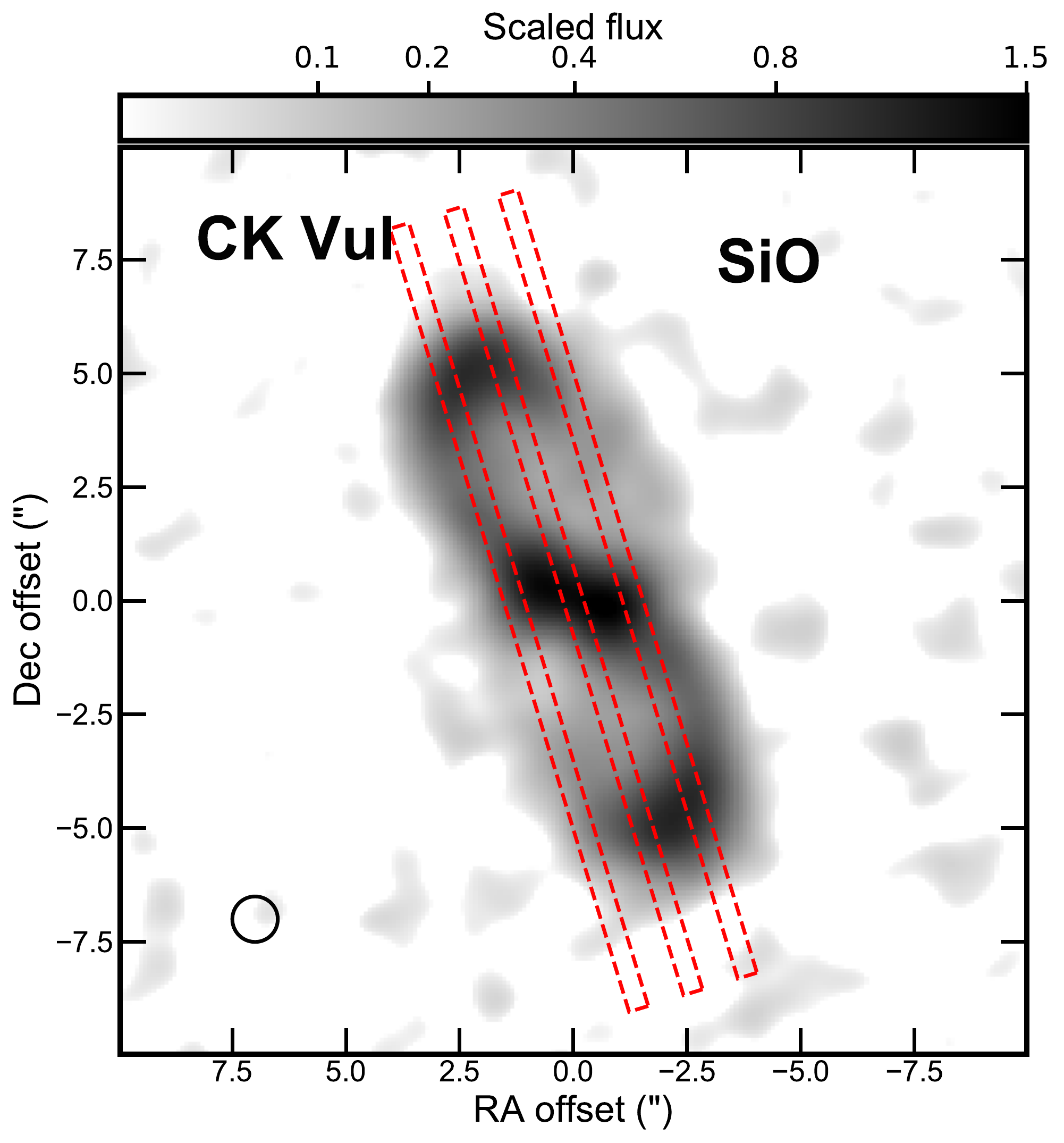}
  \includegraphics[width=.48\columnwidth]{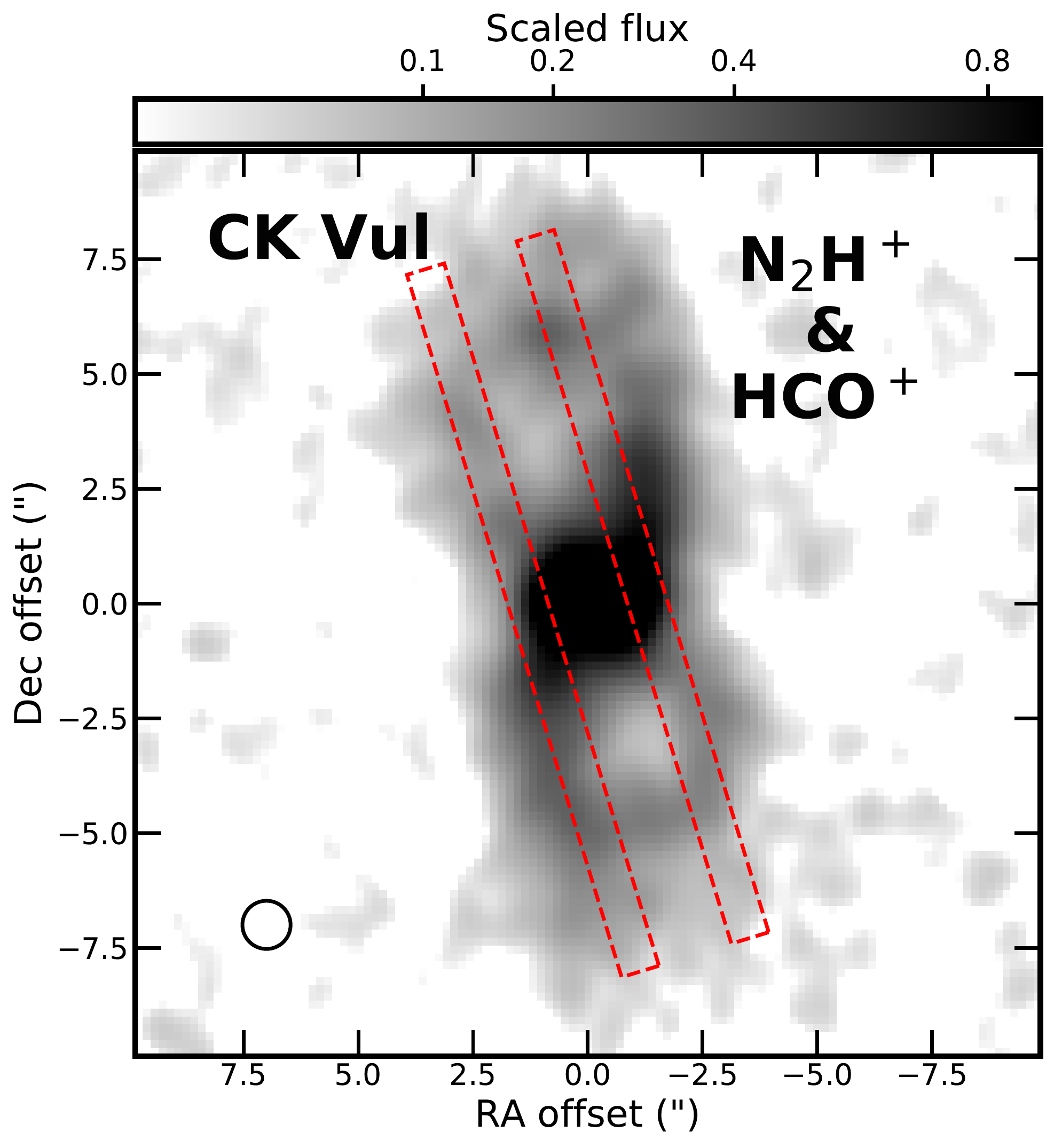}
\caption{Locations of virtual slits used to generate the position-velocity diagrams in Figs.\,\ref{fig-PV-neutrals} and \ref{fig-PV-ions}. The background images show weighted-mean emission in several transitions of SiO (left) and combined N$_2$H$^+$ and HCO$^+$ emission (right) (cf. Paper\,I). }\label{fig-slits-ions}
\end{figure}

\begin{figure*}
  \includegraphics[width=.33\textwidth]{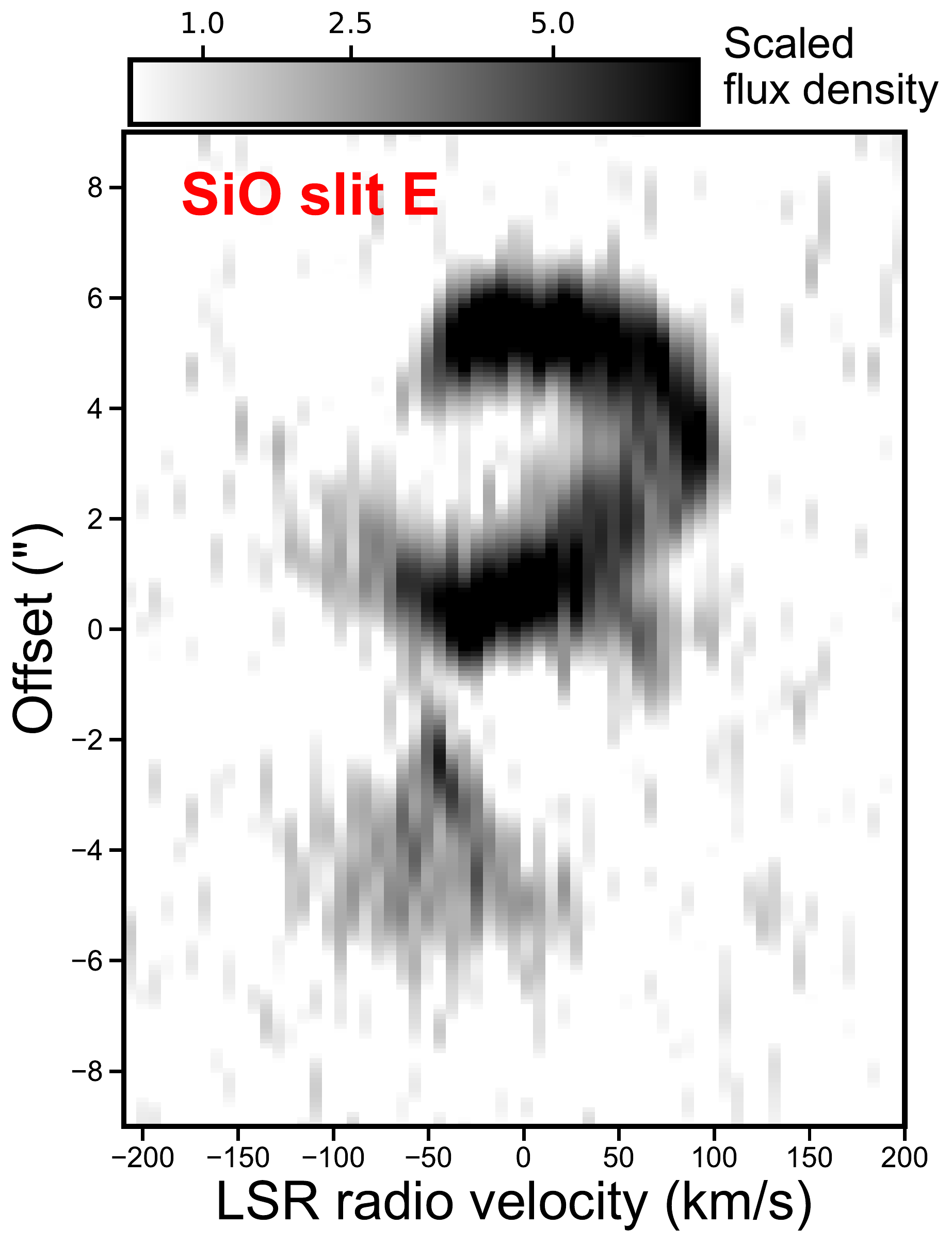}
  \includegraphics[width=.33\textwidth]{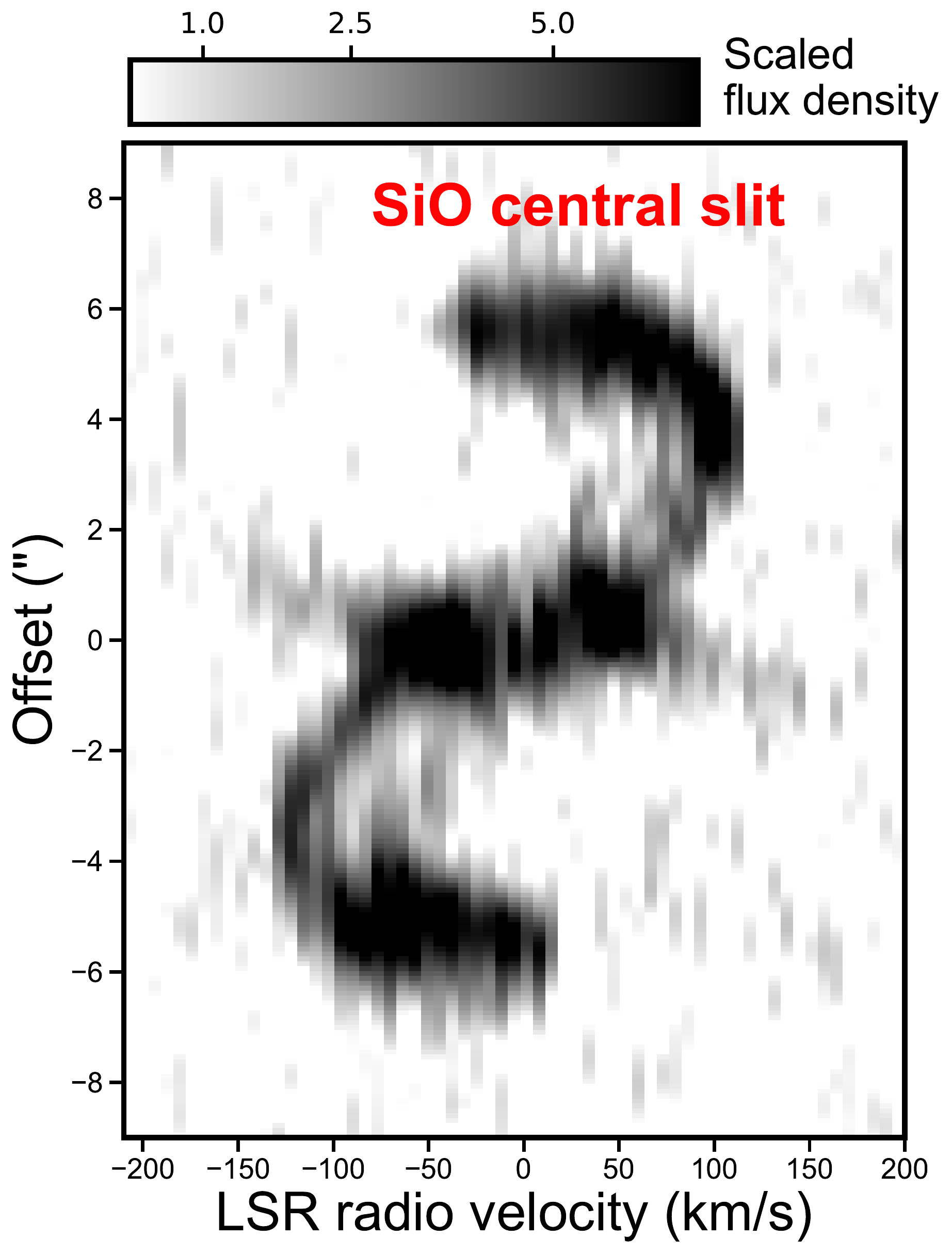}
  \includegraphics[width=.33\textwidth]{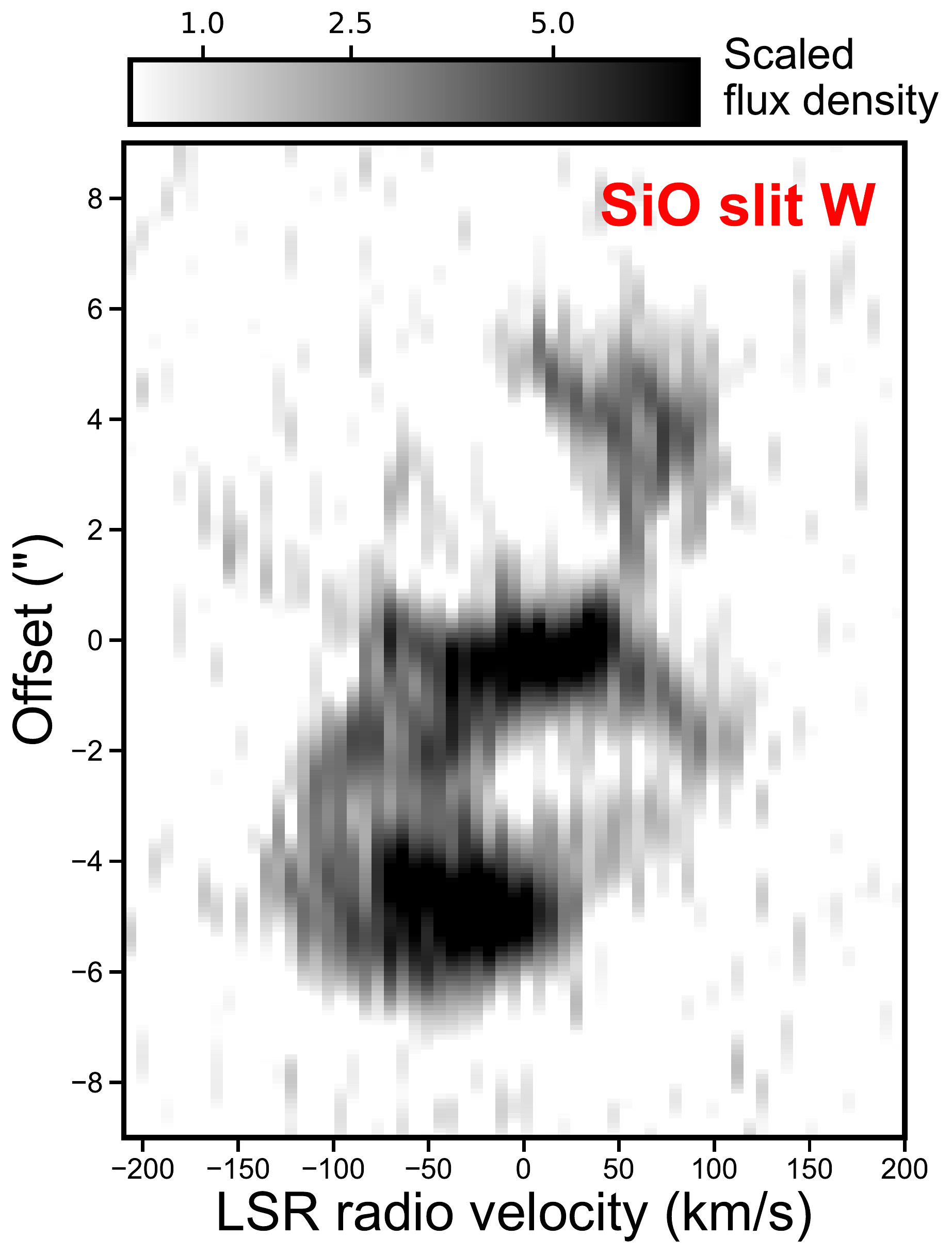}
  \includegraphics[width=.33\textwidth]{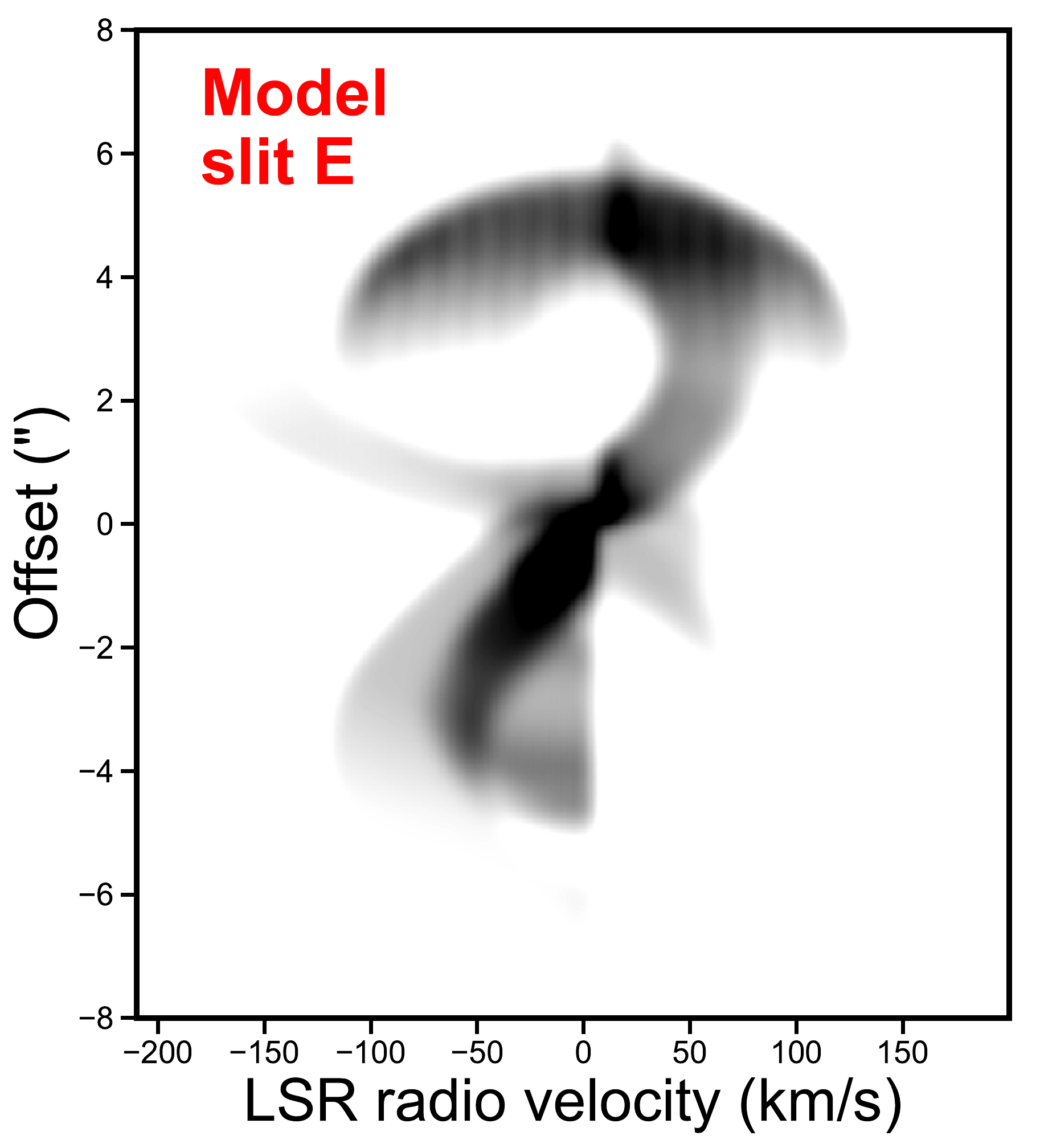}
  \includegraphics[width=.33\textwidth]{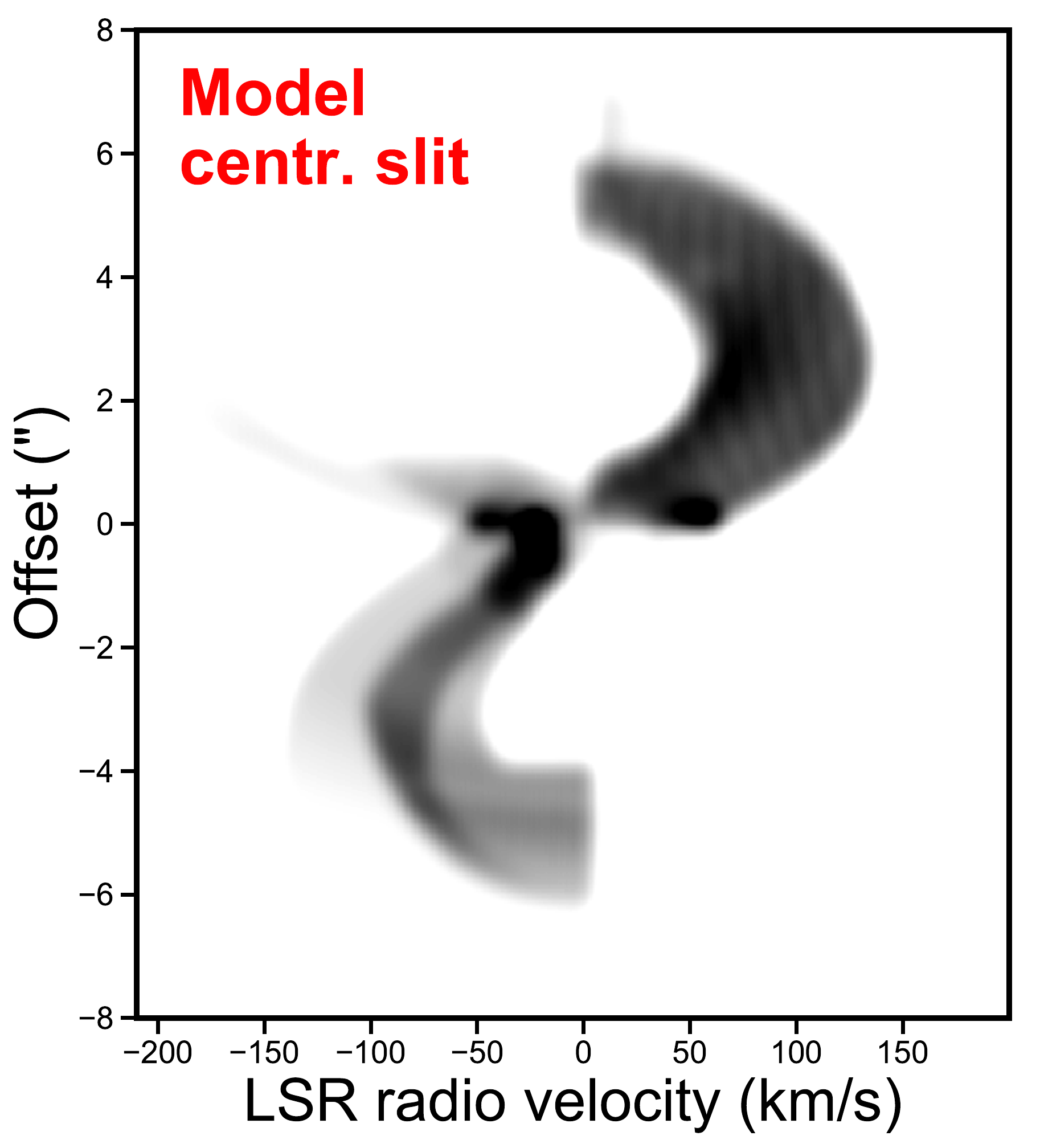}
  \includegraphics[width=.33\textwidth]{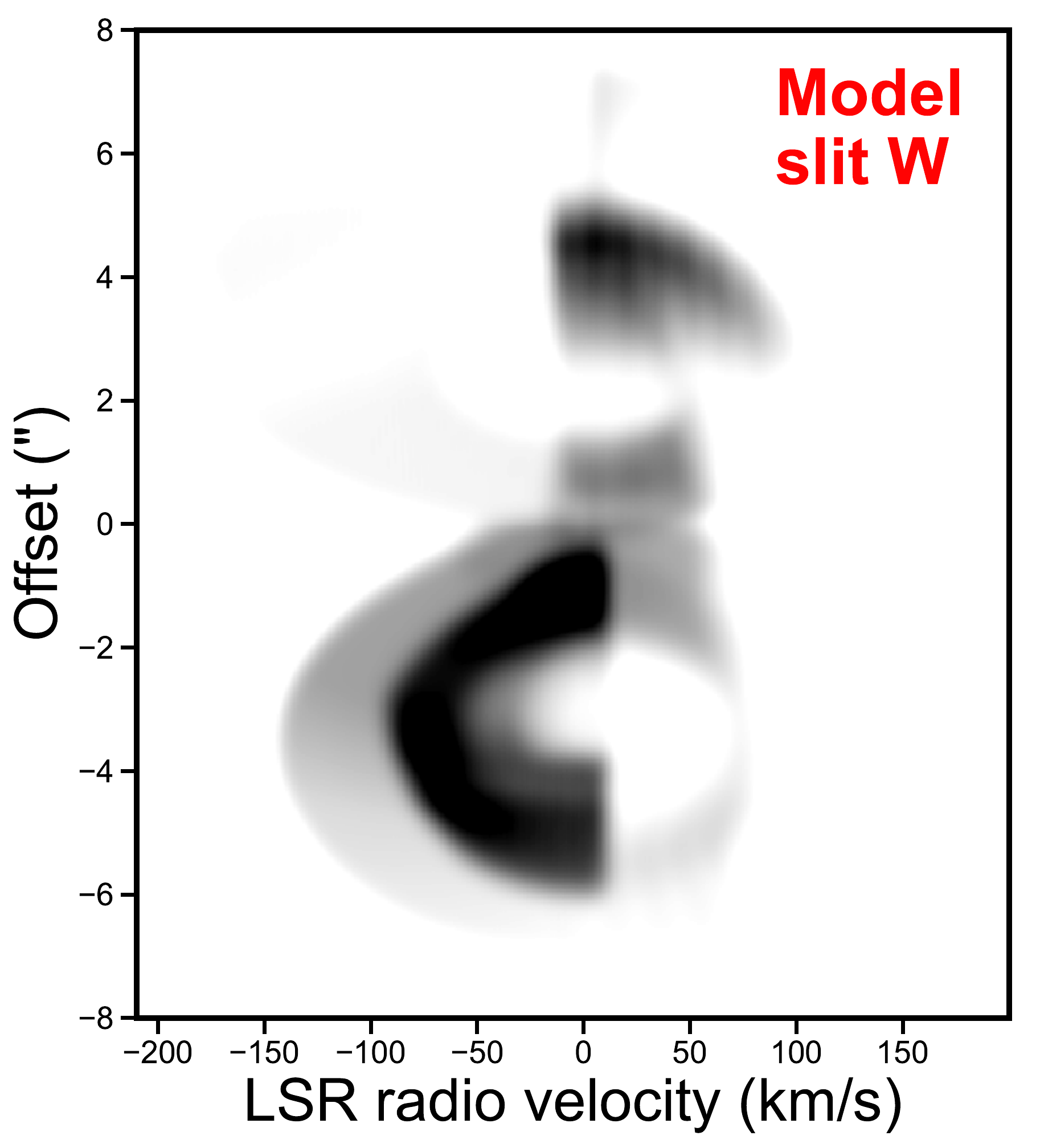}
\caption{Comparison of observed (top) and modeled (bottom) position-velocity diagrams of SiO emission along the slits shown in Fig.\,\ref{fig-slits-ions}.}\label{fig-PV-neutrals}
\end{figure*}

We filled the shells with virtual gas whose density was parametrized in spherical coordinates ($r, \theta, \phi$). In Paper\,I we found that the lobes can be characterized with a single temperature of about 14\,K. Then, in the approximation of optically thin emission, all relative intensity variations are related to changes in the local density of the molecular tracer. In the SHAPE mode that we used, it is possible to define density changes as a function of one of the spherical coordinates but it is not straightforward to implement density variations simultaneously in two or more coordinates.
This somewhat limited our flexibility in reproducing the observations, but still provided us with constraints on the gas distribution. 
Since no realistic radiative transfer calculations were performed, we were aiming for only a qualitative reproduction of the intensity maps. The PV diagrams of many neutral species with extended emission required limiting the distribution of matter to certain ranges of the azimuthal angle, $\theta$. (As indicated in Fig.\,\ref{fig-shape1}, the angle $\theta$ is defined relative to the sky plane and counted positive from the west toward the observer.) For instance, in the northern lobe nearly no emission is assigned to $10\degr < \theta < 90\degr$ (i.e., the western part of the near side is empty) and our modeled southern lobe is devoid of emission for $5\degr < \theta < 185\degr$ (i.e., nearly the entire far side is missing). Although it is certain that these density variations with $\theta$ exist, their magnitude and the exact limits on $\theta$ could not be well constrained. 
A comparison between sample PV diagrams of simulated and observed emission of SiO is shown in Fig.\,\ref{fig-PV-neutrals}. These diagrams correspond to three virtual slits defined in Fig.\,\ref{fig-slits-ions}.


\begin{figure}\centering
  \includegraphics[width=.99\columnwidth]{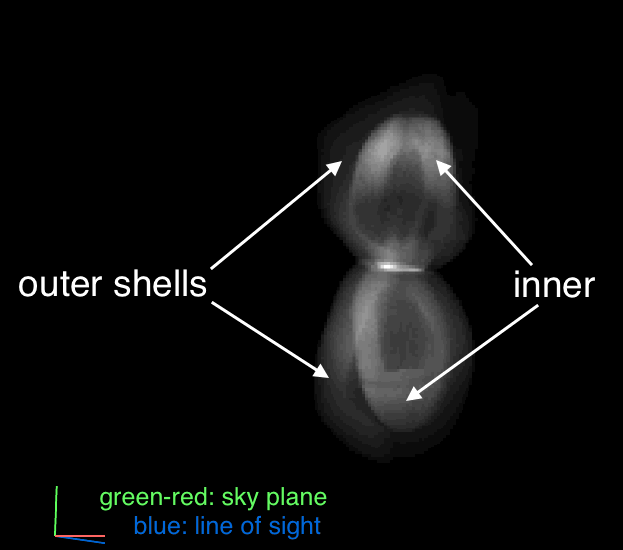}
\caption{Animation presenting the reconstructed density structure of the neutral gas in CK\,Vul. [Here will be an A\&A-compatible animation in .mov format; for now, see {\tt{https://bit.ly/3hR6jSF}}; the same for ions: {\tt{https://shorturl.at/gwzOT}}.] The blue bar corresponds to the line of sight.}\label{fig-animation-neutrals}
\end{figure}

Our model of neutral gas was arbitrarily implemented starting from elliptical shells which were devoid of gas for a range of azimuthal angles. However, alternatively the structure could have been implemented as an ensemble of streams or jets injected into the larger shell at discrete azimuthal and inclination angles. These streams or jets would have to be injected close to the walls of the larger shell to result in the S-shaped morphology. A single jet, even a precessing one, is highly unlikely to be responsible for the observed characteristics. 

\subsection{Step 3: Molecular ions}\label{sec-model-ions}
\begin{figure}\centering 
  \includegraphics[width=\columnwidth, trim=45 0 45 0, clip ]{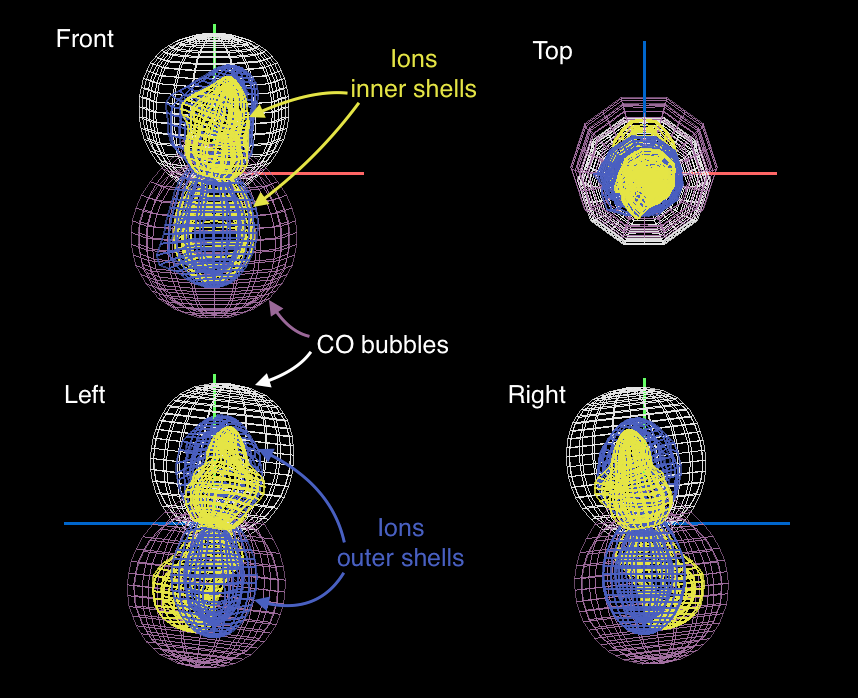}
\caption{Combined model of nested shells. The model includes two large-scale CO bubbles (pink and white) and shells of ionized molecules (blue and yellow). The outer shells from Fig.\,\ref{fig-shape1} are both shown here in blue.}\label{fig-bubbles}
\end{figure}

\begin{figure}
  \includegraphics[width=.49\columnwidth]{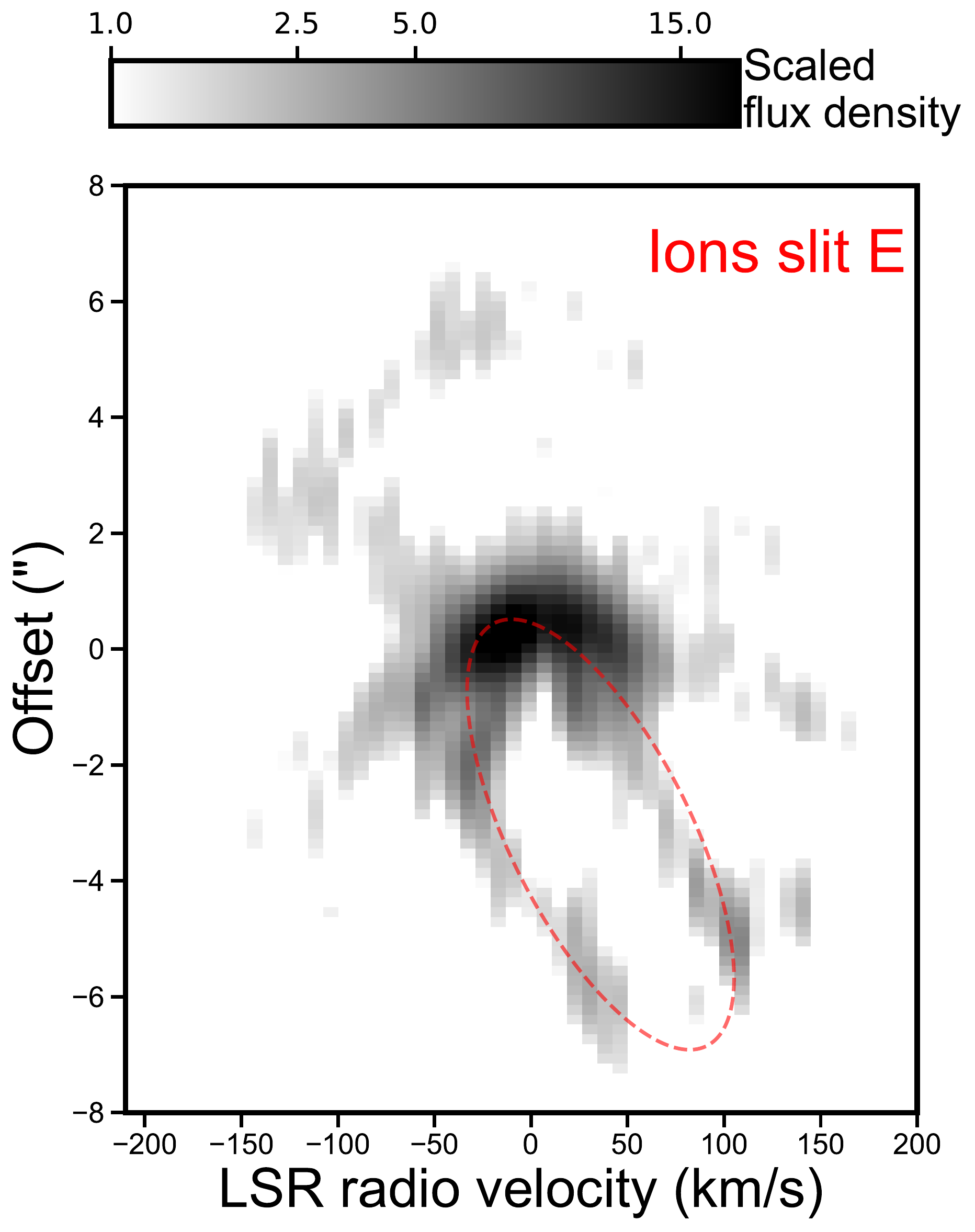}
  \includegraphics[width=.49\columnwidth]{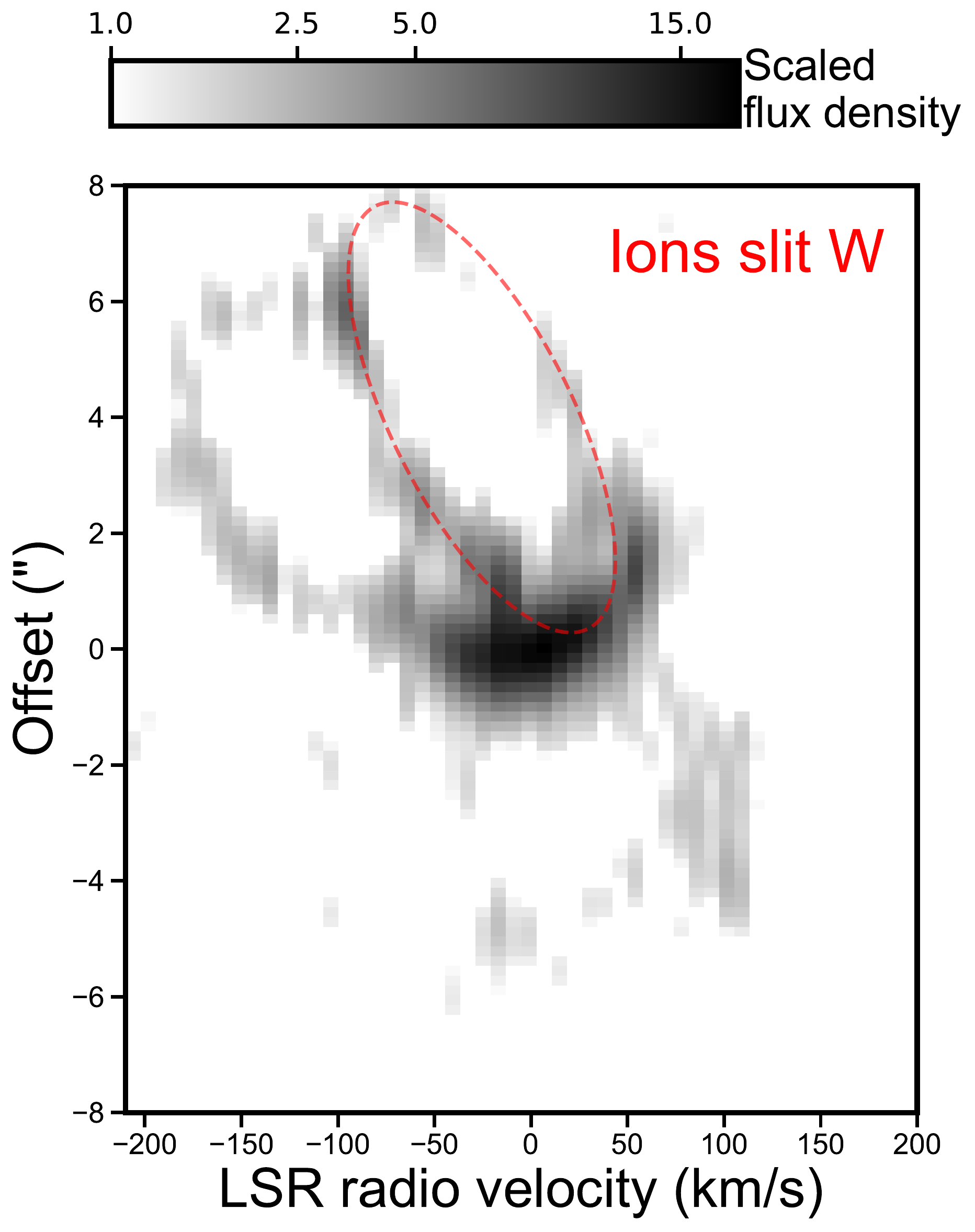}
  \includegraphics[width=.49\columnwidth]{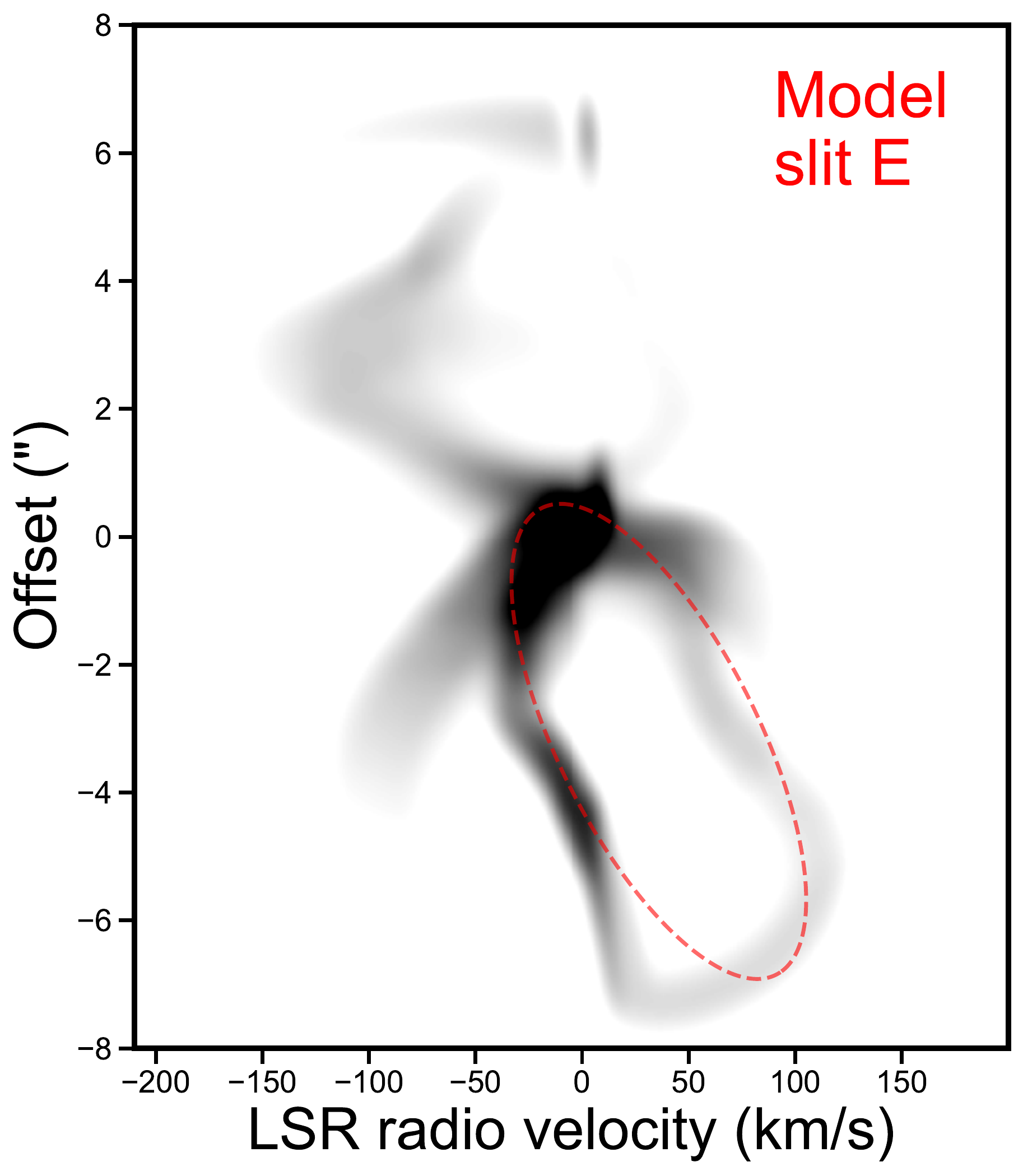}
  \includegraphics[width=.49\columnwidth]{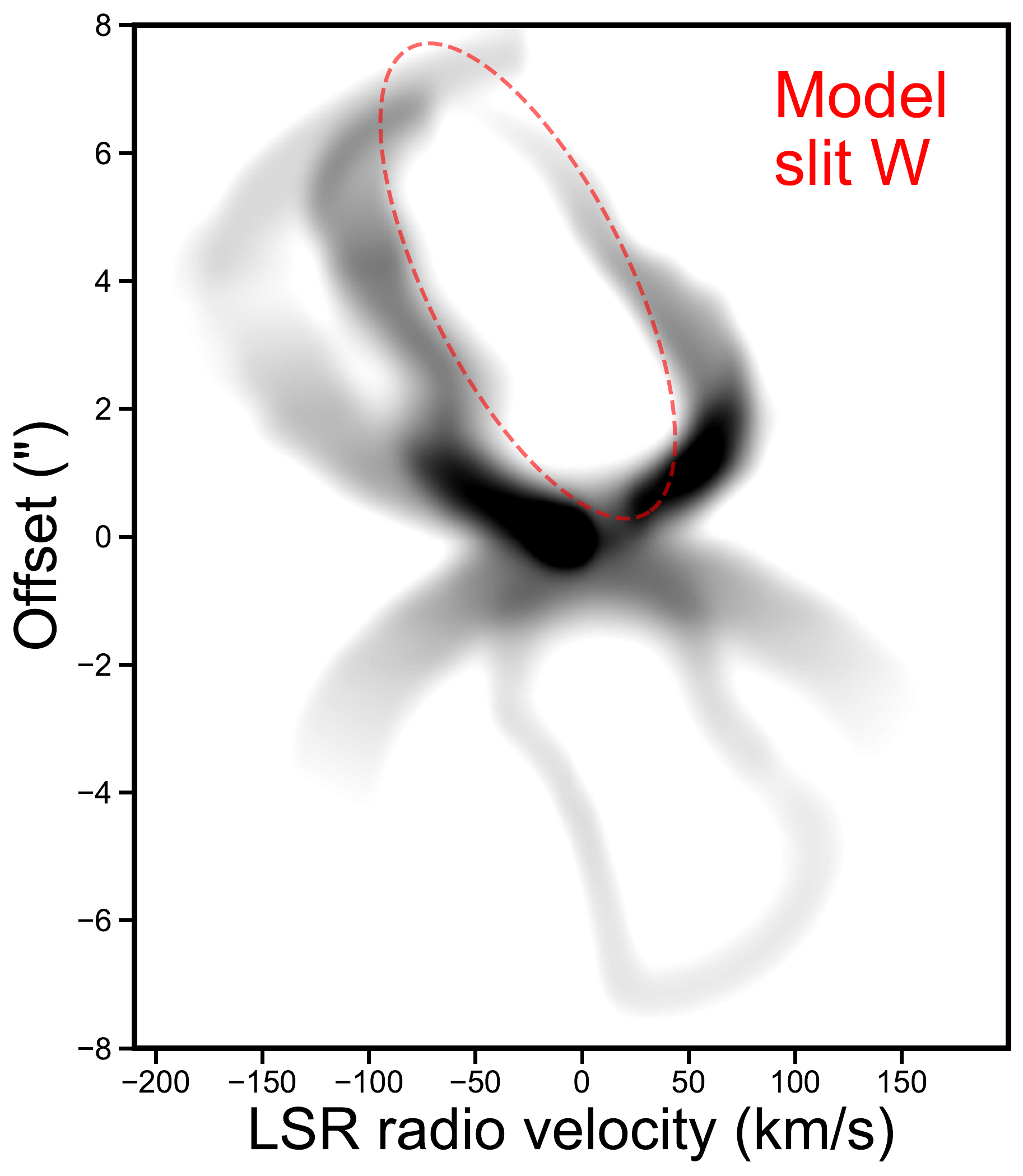}
\caption{Comparison of observed (top) and modeled (bottom) PV diagrams of ionic emission along the slits shown in Fig.\,\ref{fig-slits-ions}. Ellipses drawn with red dashed lines mark locations of loops discussed in the text.} \label{fig-PV-ions}
\end{figure}

In Paper\,I, we showed that ions display a different spatial distribution than most neutral molecules. Our 3D analysis of the emission confirms that notion. We analyzed the extended emission of molecular ions in 3D based on a data-cube representing the combined emission of several transitions of HCO$^+$ and N$_2$H$^+$ (cf. Paper\,I). In constructing the 3D model, we focused on qualitatively reproducing the most characteristic features, but we ignored the patchy appearance of the emission. Our preferred model is shown in Fig.\,\ref{fig-bubbles} and it is compared to observed PV diagrams in Fig.\,\ref{fig-PV-ions}. The diagrams were extracted for virtual slits shown in Fig.\,\ref{fig-slits-ions}. 

The thin outer shell seen in most neutral molecules (and shown in Fig.\,\ref{fig-shape1}) is seen in ions too, but is very incomplete. Its emission is very weak at large distances from the center and is almost absent for several ranges of the azimuthal angle (e.g., between 70\degr\ and 180\degr\ in the northern lobe). Our model replicates only very roughly the irregularity of the weak outer shell. 

Molecular ions display strong features in the PV diagrams that cannot be reproduced by our basic model for the emission from neutral molecules. 
These are outlined by  red dashed ellipses in Fig.\,\ref{fig-PV-ions}. We reproduced these features seen in emission from cations by introducing a couple of incomplete shells embedded within the outer shells, one in each lobe. They are scaled-down versions of the outer shells but with multiple shape modifications (bumps) added, which resulted in very irregular structures, as shown in Fig.\,\ref{fig-bubbles}. 

As before, a homogeneous linear velocity field ($\varv\propto r$) at low inclination angles ($i$=5\degr\ or 10\degr) was assumed for the gas from ionic species. The gas assigned in the simulations to the inner shells generally moves slower than that in the outer shells. Our adopted linear velocity fields imply kinematic ages of 240--470\,yr (at 3.5\,kpc). Within the model uncertainties, the age is thus consistent with origin in the 17th-century eruption. Constructing a model that matches exactly the age of the remnant is possible, provided more effort and more advanced modeling tools. 

In combination with the very complex shape of the shells, the linear velocity field satisfactorily reproduces most of the observations. Alternatively, introducing more regular shell structures, such as a couple of pure ellipsoids, would require a very complex, highly non-linear, velocity field which we consider unlikely. We also experimented with an alternative model where each ionized lobe is represented by 3--4 simple ellipsoidal shells of different sizes, inclinations, and (different) linear velocity fields. That model was satisfactorily reproducing the observations but requires clearing big pars of those shells (cf. below). The complex spatio-kinematic characteristics of ionic emission does not warrant a unique 3D model.


An analysis of the distribution of ionized gas within the shells also indicates complex structures. To reproduce the observed PV diagrams, inner shells were filled with emitting matter only for very narrow ranges of the azimuthal angle. For instance, for the northern lobe the emitting matter was located within 55\degr$<\theta<$80\degr\ and --80\degr$<\theta<$0\degr. These ranges correspond, respectively, to the mainly redshifted and mainly blueshifted parts of the loops seen in the PV diagrams of the western side. The narrow limits on azimuthal angles indicate a very confined location of molecular ions. The features they display in  the  northern lobe can alternatively be well reproduced with a pair of collimated jet-like structures extending along and inside the walls of the outer shell. 



The features producing the loops in Fig.\,\ref{fig-PV-ions} are different from the bulk of the material traced in neutral molecules, but their origin and relation to other components of the remnant are unclear. Ions may represent a more turbulent phase of the circumstellar gas than that probed by neutral molecules. One might ask, whether these ionized jet-like structures are younger (e.g., more recently shocked) than those seen in emission of neutrals? Neither molecule formation nor the dynamics of these features are understood well enough to provide a satisfactory answer. 

\subsection{Step 4: Extended bubbles of CO emission}\label{sec-bubbles}
\begin{figure}
  \includegraphics[width=.99\columnwidth]{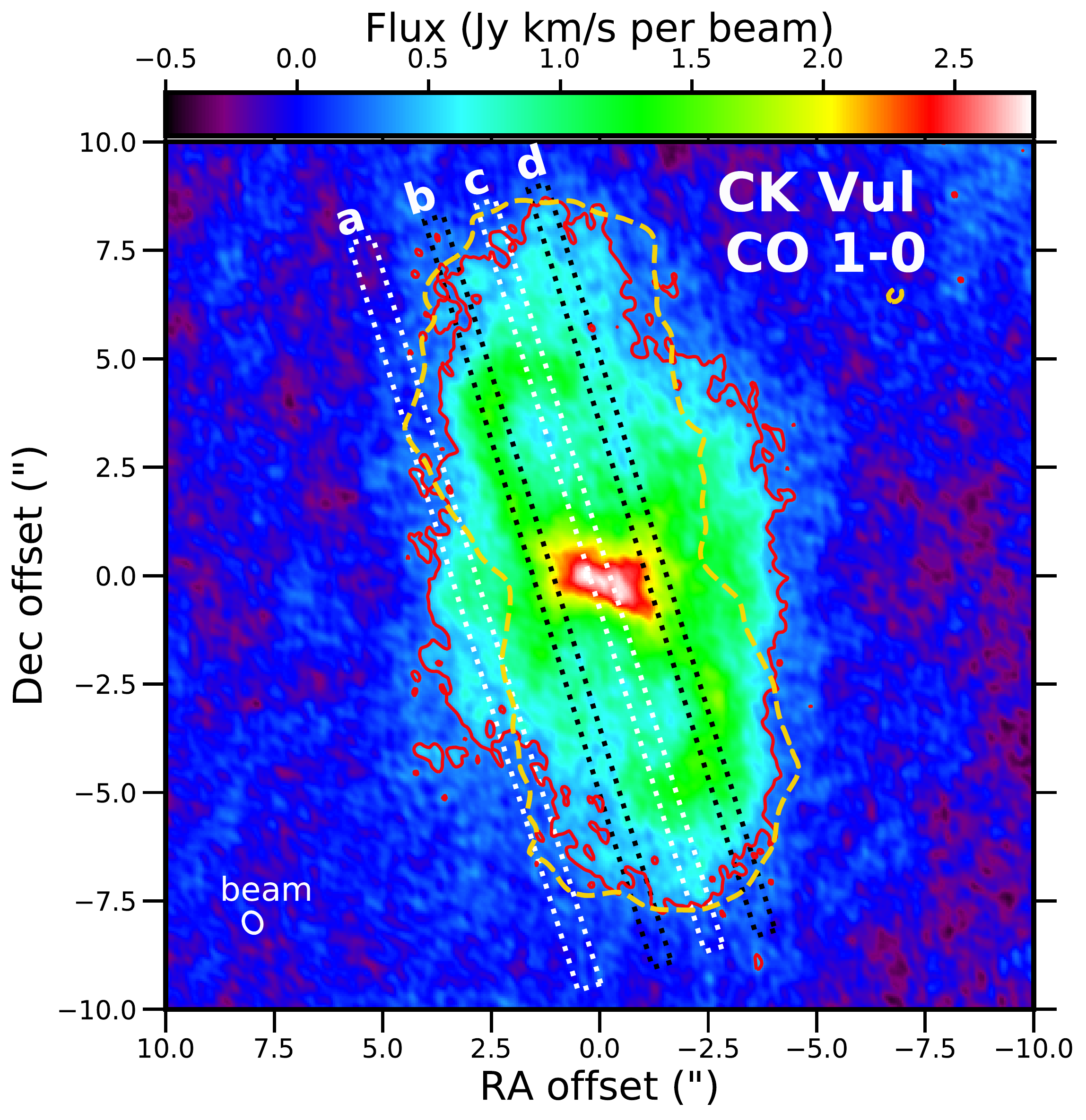}
\caption{Total intensity map of $^{12}$CO 1--0 restored with natural weighting of visibilities. The 3$\sigma$ level is shown with a red contour. The orange dashed contour shows the 3$\sigma$ level for combined emission in several lines of HCN and H$^{13}$CN. It illustrates a much larger extent of CO compared to HCN (and all other molecules) near the central waist. Dashed lines indicate locations of four virtual slit for which PV diagrams are shown in Fig.\,\ref{fig-co-PVs}.}\label{fig-co10}
\end{figure}

Emission observed in the 1--0 and 2--1 lines of CO has a component that is more extended than any other molecular feature observed with ALMA and SMA. Most apparent is the extra emission east and west from the central waist of the bipolar structure, as illustrated in Fig.\,\ref{fig-co10}. The CO gas is partially outside of the outer shell discussed above and shown in Fig.\,\ref{fig-shape1}. The extended component is patchy in the outermost parts, but its overall distribution has a pronounced point symmetry. 
The CO emission may even be more extended than it appears in our ALMA maps because the interferometric data are not sensitive to structures larger than $\approx$23\arcsec\ owing to the limited range of short interferometric baselines (Paper\,I). Indeed, compared to a single-dish spectrum of CO 1--0 obtained with a beam of 21\farcs8 \citep{kami-singledish}, the ALMA maps recover only about 52\% of the line flux. A component extending beyond the CO emission seen in ALMA maps was recovered in combined SMA and APEX data in CO 3--2 \citep{nature}. Given the low excitation temperature of CO of $\approx$10\,K (Paper\,I), the emission in CO 1--0 may extend beyond the $J$=3--2 emitting region. On the other hand, our ALMA maps recover extended emission ($\lesssim$32\arcsec) of interstellar molecular clouds in the field of view centered on CK\,Vul, which is much larger than the extent of the remnant in the same transitions. In the analysis below, we ignore the missing flux and focus on the recovered $J$=1--0 emission of $^{12}$CO and $^{13}$CO seen by ALMA. 

\begin{figure*}\centering
  \includegraphics[width=.24\textwidth]{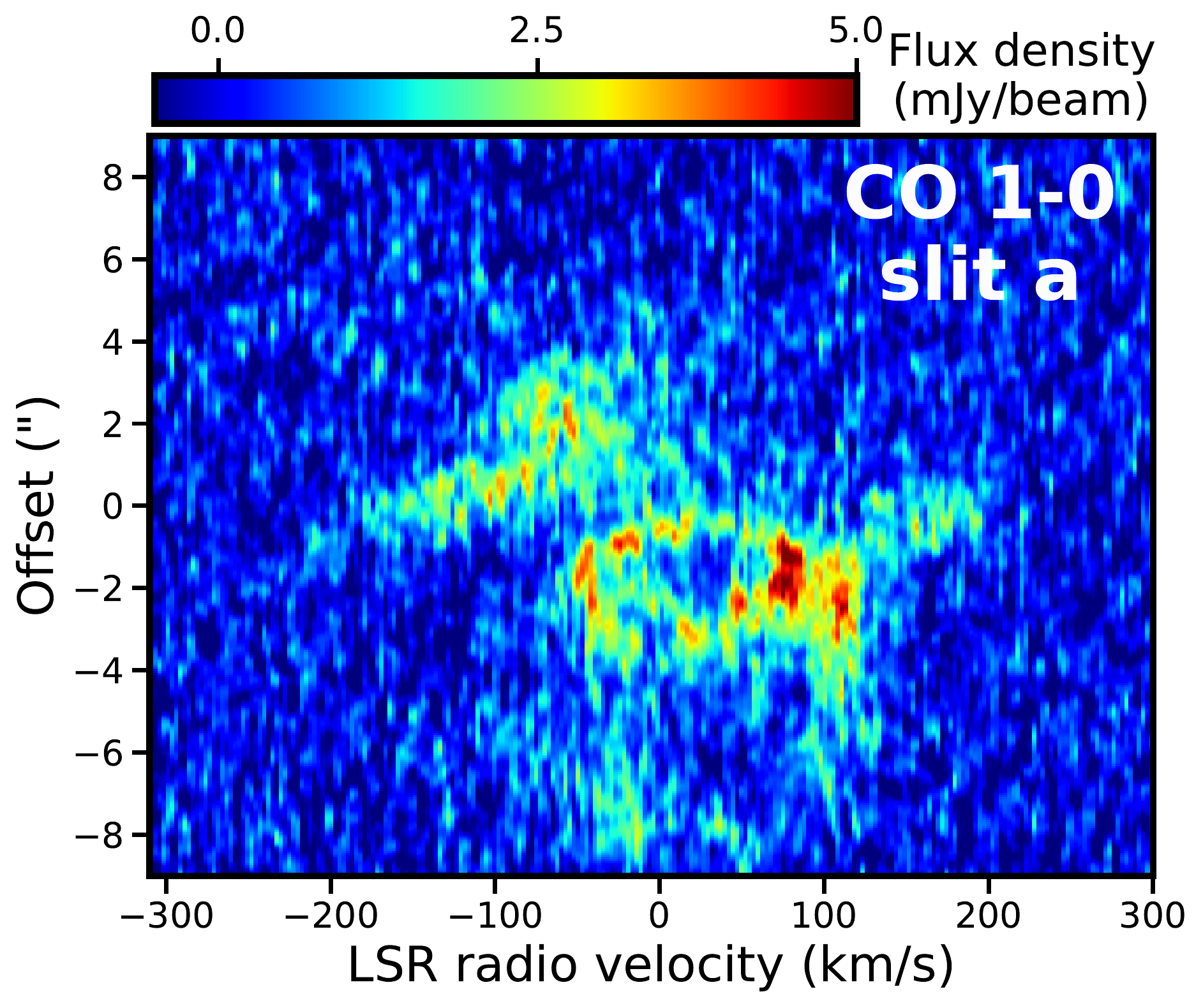}
  \includegraphics[width=.24\textwidth]{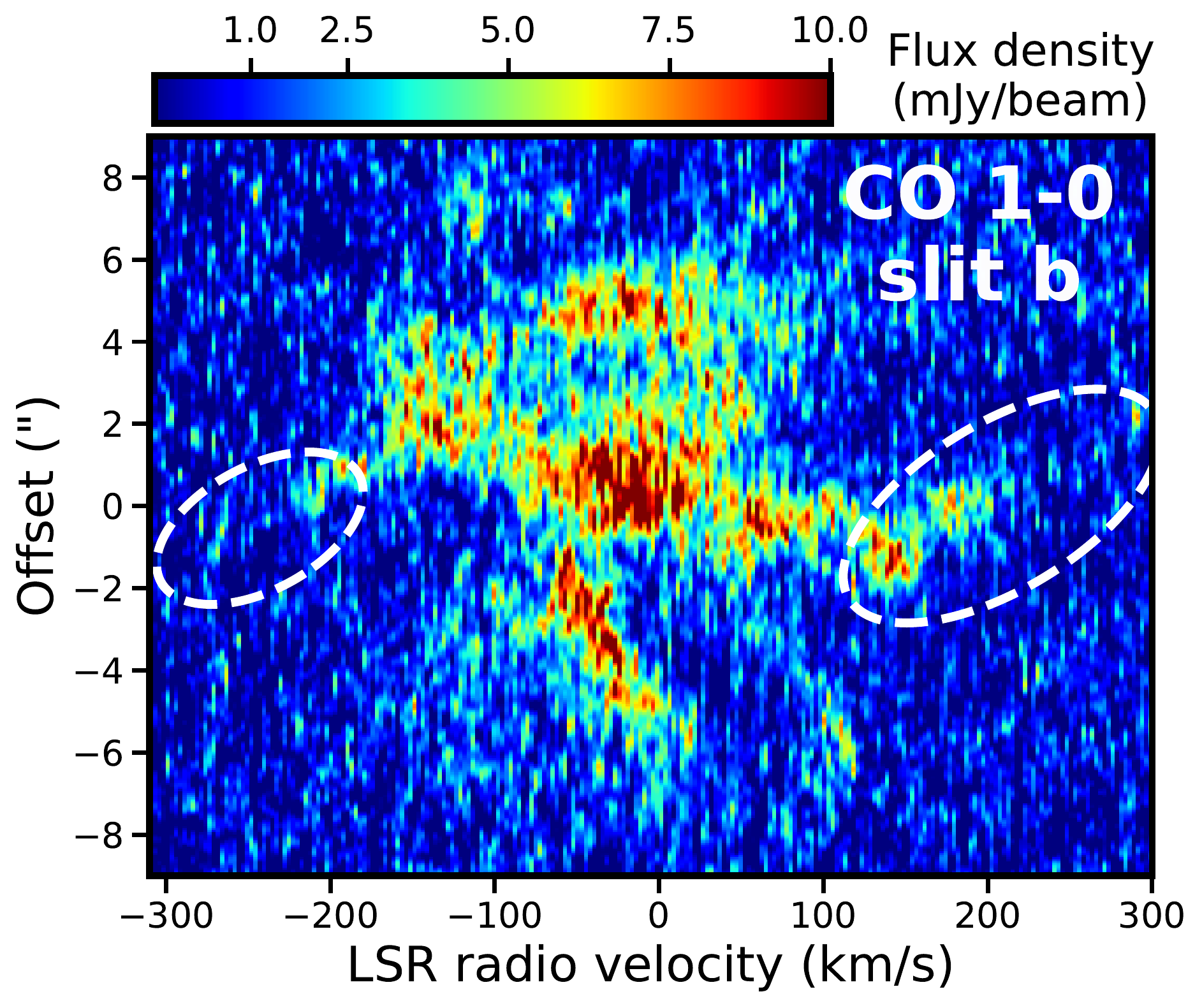}
  \includegraphics[width=.24\textwidth]{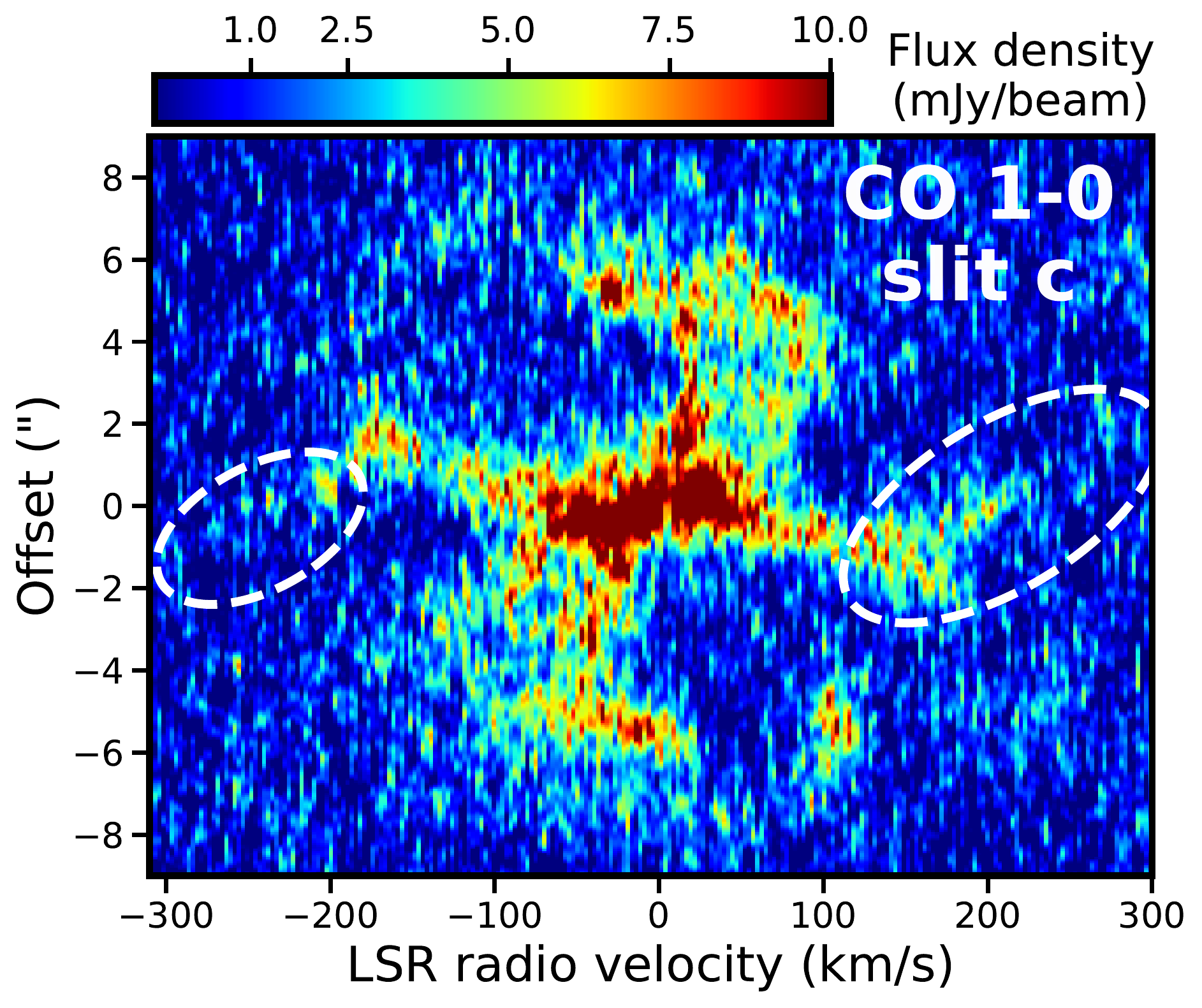}
  \includegraphics[width=.24\textwidth]{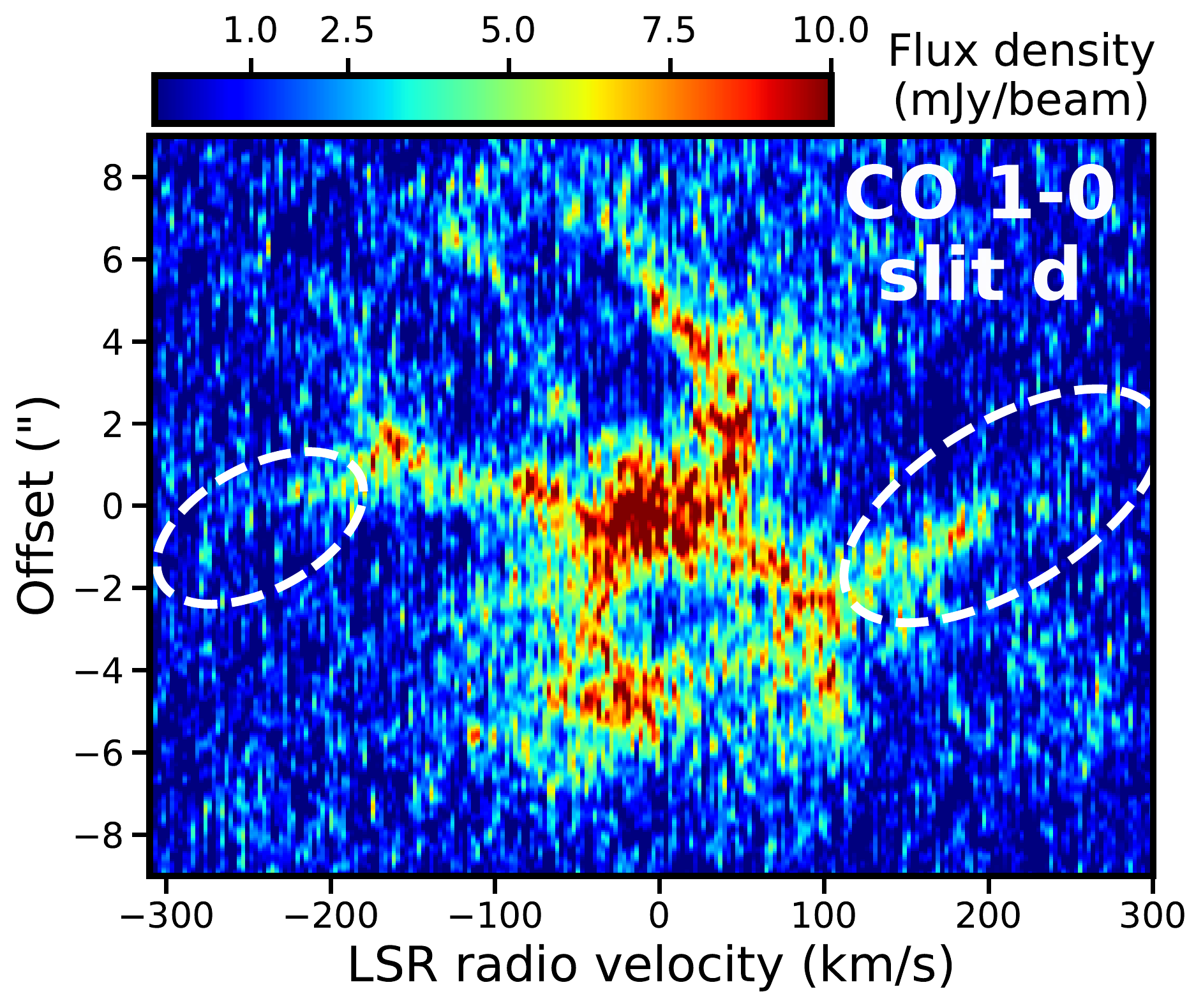}
\caption{Position-velocity diagrams of CO 1--0 emission for four virtual slits shown in Fig.\,\ref{fig-co10}. The dashed ellipses mark features which are only seen in CO emission. Entire emission seen within slit $a$ is seen exclusively in CO.}\label{fig-co-PVs}
\end{figure*}

The extended component has a very high radial velocity that implies radial motions $\gtrsim$300\,\kms\ relative to the mean LSR velocity of --10\,\kms. This fast component is shown in a sample of PV diagrams in Fig.\,\ref{fig-co-PVs}. Its high radial velocity allows us to identify the CO component also at positions close to the center of the remnant where multiple components spatially overlap in total-intensity maps. 

We were unable to construct a regular 3D representation of this extended CO structure that would fully reproduce its spatio-kinematic structure. However, an idealized model consisting of two spherical shells with a size of 16\arcsec\ explains some of the observed characteristics. In that implementation, we used a simple velocity law with $k$=71\,\kms arcsec$^{-1}$. It is the steepest function $\varv(r)$ among all analyzed features. The spheres overlap in the central 1\farcs7 part of the remnant. The shells are centered at $r$=4\farcs3 north and south from the remnant center. These CO bubbles are shown in Fig.\,\ref{fig-bubbles} as the most extended features (in white and pink). From observations, however, it is not clear how far north and south the emission extends. Also, the spherical shells are anti-symmetrically filled with material as the function of $\theta$, as evidenced by the PV diagrams. The described CO features are somewhat similar to structures recently discovered in CO emission within the PPN OH231.8+4.2 and named ``fish bowls'' \citep{SC18}.  

\begin{figure*}\sidecaption
  \includegraphics[width=12cm]{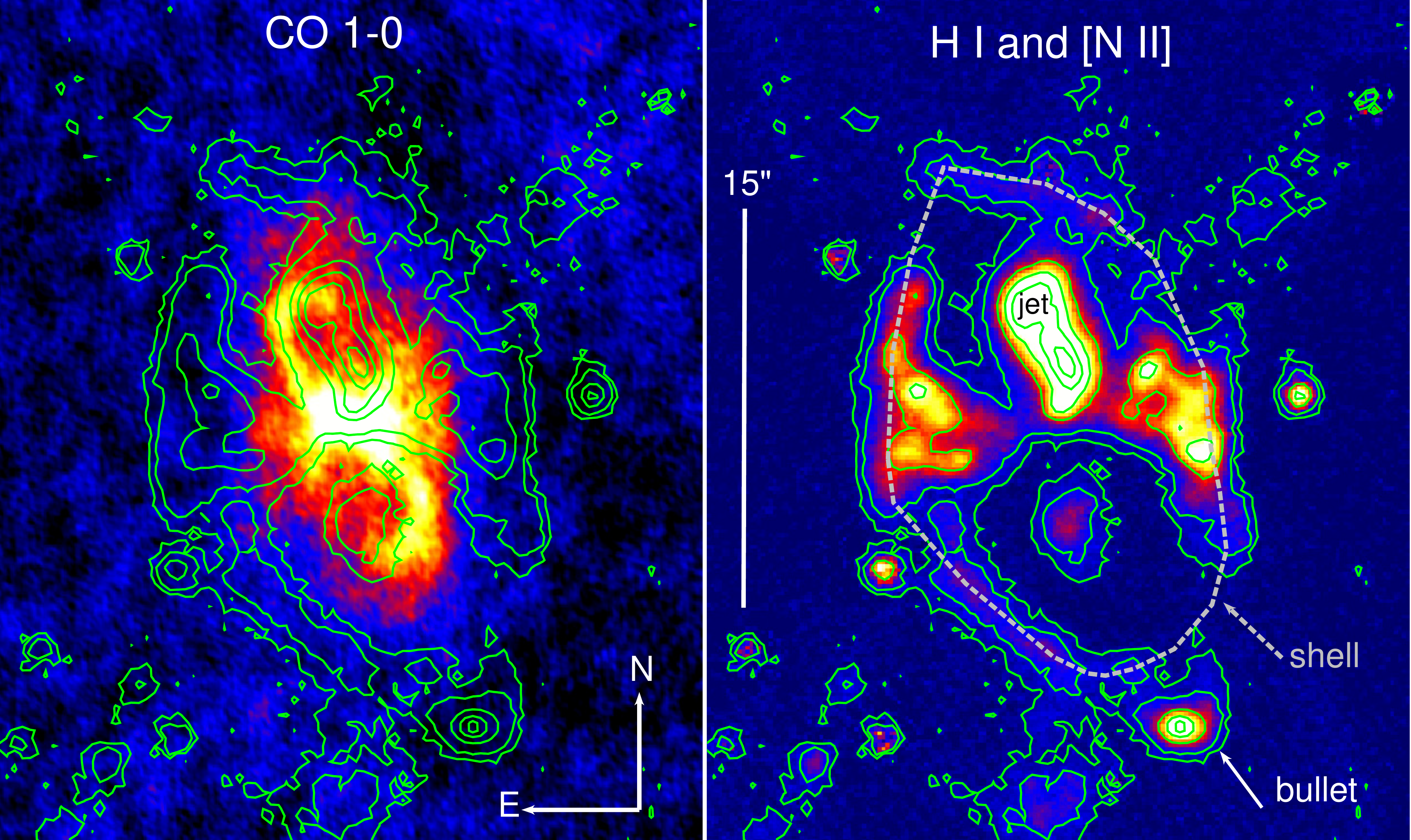}
\caption{Relative location of atomic and molecular gas in the inner remnant. The color images in the left and right panels represent, respectively,  CO 1--0 and optical nebular emission near H$\alpha$. The latter is also shown as contours on both maps. Nearly all unrelated continuum sources were removed from both maps. The main features are labeled in the right map. The dashed line roughly marks emission of a partially-broken shell of recombining gas that appears to surround the molecular remnant (see text). The bullet is one of the two features studied in \citet{hajduk2007} and whose proper motions indicate an ejection in 1670.}\label{fig-co-opt}
\end{figure*}

Some clues on the structure of the extended molecular emission come from its comparison to atomic emission seen in projection in the same part of the remnant. A comparison is made in Fig.\,\ref{fig-co-opt}, where we used a map of H$\alpha$ and [\ion{N}{II}] emission obtained in 2010 with Gemini by \citet{hajduk2013}. The 8\,yr time span between the ALMA and Gemini observations is too short to produce any changes related to tangential motions, since we are limited by our angular resolutions of 0\farcs45 in CO and 1\farcs1 in the optical.  

The elongated optical filament marked as jet in Fig.\,\ref{fig-co-opt} closely fills the gap in the loop of CO emission seen in the northern lobe. This suggests a direct link between the two phases of the remnant. The nebular emission within the southern lobe is much less extended and very weak but still constitutes a natural extension of a broken loop of CO emission of that lobe. Weak atomic emission is seen surrounding the entire CO emission region, as outlined in Fig.\,\ref{fig-co-opt}. Not counting the jet, the brightest clumps of the atomic nebula are seen 2\farcs3--7\farcs5 east and west from the center. A much weaker structure is seen in almost all directions forming a bi-conical limb-brightened structure surrounding the molecular remnant. Its main axis is at a position angle of $\sim$25\degr, close to that of the molecular nebula (17\degr). Its northern tip appears sharply pointed whereas the apex of the southern lobe is rounded and incomplete, or appears so only due to an overlap with a bullet. The shell of recombining gas surrounding the molecular remnant is very likely physically linked to the inner remnant traced in molecular emission. It was most likely created, or excited, by activity in the inner regions and may represent gas that was ejected earlier than that seen in the molecular emission and in the northern optical jet. Indeed, the kinematical age of the bullet agrees with an ejection near 1670 \citep{hajduk2007}. Since the low kinetic temperature of the molecular gas ($\approx$10--14\,K) requires a phase of strong adiabatic cooling, the atomic shell may contain material on which work has been done by the expanding molecular gas.     


\section{Mass, momentum, and kinetic energy in the lobes}\label{sec-ekin}
As mentioned in Sect.\,\ref{intro}, PPNe have the peculiar characteristics that the linear momentum and the kinetic energy of their bipolar outflows exceed by orders of magnitude what the central star is capable of releasing as radiation over the lifetime of the nebula \citep{Bujarrabal2001}. Here we show that the remnant of CK\,Vul stores a comparably high momentum and kinetic energy. 

We follow closely the definitions of linear angular momentum and kinetic energy in the outflow of \citet{Bujarrabal2001}, who systematically characterized a large number of PPNe and related objects. In particular, we are interested here in scalar angular momentum, which represents the sum of momenta in the two opposing lobes (see \citet{Bujarrabal2001} for a discussion of the shortcomings of using this quantity). In our calculations we use results from Paper\,I where column densities of CO have been estimated in four parts of the CK\,Vul's remnant: in regions SW, NE, NW, and SE within the lobes, and in the central region. These were found with non-LTE models fit to several rotational transitions of $^{12}$CO and $^{13}$CO. By multiplying these column densities by the respective sizes of the regions, we estimated the number of CO molecules, which we next converted to H$_2$ mass using a CO abundance of 2$\times$10$^{-4}$ relative to H$_2$ \citep[as in][]{Bujarrabal2001}. We consider three alternative distances to the source when calculating the physical areas of the regions: the lower limit of 2.6\,kpc, the most likely value of 3.5\,kpc, and our working maximal value of 5.7\,kpc (Sect.\,\ref{sec-hourglass}). The column densities and masses are listed in Table\,\ref{tab-momenta}. The masses are underestimated for several reasons. ({\it i}) The $^{12}$CO lines are optically thick in some regions and our radiative transfer procedure applied only a rough correction for the line saturation. ({\it ii}) About 50\% of flux is missing in ALMA maps owing to missing ``short-spacings'' (Sect.\,\ref{sec-bubbles}). ({\it iii}) The regions considered in Paper\,I encompass 63\% of the total flux of CO 1--0 emission. There are other considerable sources of uncertainty in the calculation, including the actual abundance of CO given the peculiar chemical composition of the nebula. Hot atomic and molecular gases, although certainly present in the nebula, are not accounted for \citep[in accord with calculations in][]{Bujarrabal2001}. 

The total mass is higher than 0.3\,M$_{\sun}$ or $>$0.4\,M$_{\sun}$ when roughly corrected for the incomplete coverage of the CO emission by analyzed regions. When further corrected for the flux filtered out by the interferometer, we obtain a mass $>$0.8\,M$_{\sun}$. Most mass is stored in the central waist. The lobes constitute only 40\% of the total mass. We note that the derived masses are comparable to 1\,M$_{\sun}$ given in \citet{nature} but the latter corresponds to the invalid distance of 0.7\,kpc and is close to current estimates only due to a numeric error \citep[cf.][]{Eyres}. 

To calculate the momentum ($M\varv$) and kinetic energy ($E_{\rm kin}$), we used deprojected velocity profiles, $\varv(r)$, constrained by our 3D models of the lobes (Sect.\,\ref{sec-outer}). Using these, we calculated a flux-weighted average CO velocity in the lobes (at $r>$2\arcsec) of CO of 312\,\kms. 
The central region ($r<$2\arcsec) is occupied by slower gas, with typical velocities of $\approx$50\,\kms. 

The linear momenta and kinetic energies of the different components of the molecular remnant are listed in Table\,\ref{tab-momenta}. The lobes carry a few times more momentum (80\% of the total) than the central region and contain almost all the kinetic energy. Values at the considered distances are within the ranges of $M\varv$ and $E_{\rm kin}$ found by \citet{Bujarrabal2001} for PPNe, that is, of $10^{37}-10^{40}$\,g\,cm\,s$^{-1}$ and $10^{44}-10^{47}$\,erg, respectively. This demonstrates that the outflows of the merger remnant of Nova\,1670 are very similar to these of PPNe and thus may have a similar dynamical origin. Within the uncertainties, the kinetic energy stored in the molecular remnant of CK\,Vul is also relatively close to the value of 10$^{46}$\,erg found for two other red-nova remnants, V4332\,Sgr and V838\,Mon \citep{submmRN}.

\begin{table*}
    \caption{Characteristics of the molecular gas. (See text.)}\label{tab-momenta}
    \centering\footnotesize
    \begin{tabular}{ccc| ccc| c| ccc| ccc}
\hline    
       &$N_{\rm CO}$&Area         &\multicolumn{3}{c|}{H$_2$ mass (M$_{\sun}$)}&Velocity &\multicolumn{3}{c|}{Momentum (g\,cm\,s$^{-1}$)} &\multicolumn{3}{c}{$E_{\rm kin}$ (erg)}\\
Region &(cm$^{-2}$) &(\arcsec)$^2$&2.6\,kpc & 3.5\,kpc & 5.7\,kpc                &(\kms)   &2.6\,kpc & 3.5\,kpc & 5.7\,kpc                 & 2.6\,kpc &  3.5\,kpc & 5.7\,kpc \\
\hline\hline
SW	&2.1e17&  13.517& 0.04& 0.07& 0.17& 312& 2.2e39& 4.1e39&  1.1e40&  3.5e46& 6.3e46& 1.7e47\\
NE	&3.4e17&  11.773& 0.05& 0.09& 0.24& 312& 3.2e39& 5.7e39&  1.5e40&  4.9e46& 8.9e46& 2.4e47\\
NW	&1.1e17&  14.856& 0.02& 0.04& 0.10& 312& 1.3e39& 2.3e39&  6.2e39&  2.0e46& 3.6e46& 9.7e46\\
SE	&9.6e16&  16.898& 0.02& 0.04& 0.10& 312& 1.3e39& 2.3e39&  6.2e39&  2.0e46& 3.6e46& 9.6e46\\
Central	&1.6e18&  9.472 & 0.20& 0.35& 0.94& 50 & 1.9e39& 3.5e39&  9.3e39&  4.9e45& 8.8e45& 2.3e46\\[6pt]
All	&      &   	& 0.32& 0.59& 1.56&  & 9.9e39& 1.8e40&  4.8e40&  1.3e47& 2.3e47& 6.2e47\\
Lobes 	&      &   	& 0.13& 0.23& 0.62&  & 8.0e39& 1.4e40&  3.8e40&  1.2e47& 2.3e47& 6.0e47\\
\hline
    \end{tabular}
    \tablefoot{$N_{\rm CO}$ is the column density of CO.}
\end{table*}

\section{Discussion}\label{sec-discussion}
\subsection{Evidence of jets}  
The 3D architecture of CK\,Vul that we attempted to reproduce entails significant substructures characterized by a range of orientations (inclinations and position angles). Perhaps, the most striking is that the irregularities of the lobes are associated with a remarkable level of point symmetry between the northern and southern lobe. At least for neutral species, each bump has a corresponding feature in the other lobe, even though some appear to be more pronounced in the north. A single ejection event with a wide opening angle would have to be very asymmetric to produce such a complex shape of each lobe and at the same time preserve the strong north-south point symmetry. It is thus unlikely that a single ejection created the molecular nebula. A more natural explanation of the 3D structures is that they were formed by several collimated streams ejected at different inclination and position angles but with a high level of North-South symmetry.

The molecular remnant of CK\,Vul mimics some PPNe and Herbig-Haro objects whose morphologies are indicative of pulsed or periodic collimated outflows that take a form of jets, clumped jets, or bullets. Indeed, outflows with an S- or \reflectbox{S}-type morphology, evident in CK\,Vul, are thought to be driven by episodic jets \citep[][and references therein]{BalickFrank}. A point symmetry of such nebulae is caused by precession of objects launching them. Such jetted objects are also associated with interior bow shocks, which are present in CK\,Vul, too. Similar outflow mechanisms are very likely in action in CK Vul. Multiple streams that induce shocks naturally explain the positional variations in excitation seen across atomic regions of CK\,Vul  \citep{xshooter}; they also explain the asymmetries in the relative distribution of the neutral and ionic molecular species in CK\,Vul. 

\subsection{Changes in the lobes and their age}
Owing to multiple model degeneracies, our 3D analysis did not provide us with practical constraints on the age of the molecular lobes. The molecular lobes are however expanding fast and ALMA observations decades into the future will constrain the proper motions of its sub-components improving our understanding of the true 3D structure and age.  

The dusty and molecular nebula is changing on time scales of years and on spatial scales of the order of $\mu$as. 
this is evidenced by variability of two field stars located behind the southern lobe \citep{hajduk2013}. Because these distant stars have very small angular diameters of a few $\mu$as at the distance of CK\,Vul, even small changes in the dust distribution within the lobe can cause observable changes in their extinction. 
Since the variability is limited to two lines of sight, it does not put constraints on the age or structure of the nebula.

At much lower angular resolutions of 0\farcs6--1\farcs5, the shapes of discrete features of the optical nebula, including the northern jet, have remained unchanged since the discovery of the optical remnant in 1982 \citep{shara85}, so for at least 38\,yr. (The nebula is becoming fainter, though.) These atomic features are thus certainly older than 38\,yr but cannot be much older than their recombination time of $\approx$250\,yr \citep{xshooter}. Two bullet-like structures within the optical nebula studied by \citet{hajduk2007} have proper motions indicative of an ejection in 1670 and thus have an age of 350\,yr. It has not been clear, however, if the material responsible for the nebular emission closer to the star than the bullets was ejected at the same time and in the same event as the bullets and as the large hourglass. Based on the correspondence in the distribution of optical and CO 1--0 emission discussed in Sect.\,\ref{sec-bubbles}, we postulate that the origin of the shock-excited atomic features in the direct neighborhood of the molecular nebula, including the jet, is the same as that of the molecular matter. Molecules are most likely seen in the post-shock cooling zones associated with the recombining plasma behind the shock fronts (cf. Paper\,I). This sets the age of the molecular nebula between 38 and $\sim$250\,yr. We cannot exclude, however, that the molecular lobes are older (350\,yr) and formed in the latest phases of the 1670--172 eruption, when the central object became cool enough to support the production and survival  of molecules.

\subsection{Jet precession and binarity}
The remnant components can be interpreted as ejected from the central source at varying orientations. The variation in the position angle is evident in Fig.\,\ref{fig-co-opt-largescale}. The long axis of the atomic hourglass nebula is at a position angle that is $\approx$--15\degr\ off with respect to the main axis of the CO lobes. Still different is the position angle (off by $\approx$16\degr) of the atomic shell directly surrounding the molecular remnant (Sect.\,\ref{sec-bubbles}). The bullets were ejected in yet other directions. In particular, the trajectory of the south-most bullet is at $\approx$45\degr\ angle to the main axis of the CO region. Our 3D reconstruction of the neutral and ionic molecular lobes also shows evidence of reorientation of the ejection axes. 
In analogy to the PNe and PPNe, these are most naturally explained by binarity of the central stellar system. Reorientation may be related to precession in the rotation axis of one of the stars or to orbital changes in the binary induced by a circumbinary disk \citep{Artymowicz}. Depending on the true age of the molecular remnant, the ejections could have happened at the final phases of the 17th-century eruption or decades and centuries that followed. Here, we entertain the idea of a central binary system that has remained active long after the main eruption. Within the merger scenario, the system was then triple before the 1670 outburst.

Studies of irregular PPNe and symbiotic systems suggest that collimated streams are ejected by the more compact star, often during or in consequence of a periastron passage near a mass-losing companion \citep[e.g.,][]{SokerMcley, Sahai2016, KashiSoker2016}. If CK\,Vul is indeed the remnant of a stellar-merger red-nova event, it is expected to be now a bloated star with a low surface gravity and thus may act as the donor (see below). Assuming that the maximal deprojected outflow velocity of 470\,\kms\ (Sect.\,\ref{sec-outer}) is also the escape velocity of the more compact star responsible for launching of the molecular jets, we can roughly constrain the accretor's mass. For a main-sequence companion, the maximal velocity implies a mass of 0.3\,M$_{\sun}$ (a spectral type near M5) and a luminosity of 0.008\,L$_{\sun}$. Nothing that we know about CK\,Vul contradicts the existence of such a main-sequence companion, but direct observational confirmation may be beyond our current possibilities. The binarity and post-outburst accretion activity of the remnant must remain in the realm of speculations for the moment. 

We next consider the possible origin of the jet material. Based on the characteristics of other red-nova remnants, the coalesced star of CK\,Vul should resemble a red giant or a red supergiant \citep{KamiKeck, KamiSubaru, Chesneau}. Such an object could develop its own radiation-driven wind similar to those in genuine red supergiants and AGB stars \citep{Tylenda2009,Ortiz}. The material of that wind could be accreted and ejected by the postulated compact companion. Within this hypothesis, the material we see in the molecular lobes would originate directly from the wind (i.e., is not the material that had been dispersed during the merger).
If the mass-losing star managed to alter its surface chemical composition in the course of its dynamical relaxation following the coalescence, the jet material would have a different composition than the merger ejecta. We do observe chemical dissimilarities in the lobes and the central region in CK\,Vul (Paper\,I).  However, the jet material should eventually be also shocked, so the molecular composition that we observe today may not be the same as the original molecular composition of the wind. 

Alternatively, the material that is accreted and ejected by the hypothetical companion may originate from a circumbinary disk formed prior or during the 1670 merger. Such a structure may be identified in the waist of the bipolar nebula of CK\,Vul. It has been shown that in many post-AGB systems that had undergone the common envelope evolution, the circumstellar gas is transferred to and accreted onto the inner binary from a circumbinary disk that had formed in earlier phases \citep[][and references therein]{KashiSoker2016,Rafikov,OOmen2019}. It is uncertain if such a mechanism could operate in CK\,Vul, given the much shorter time scales necessary to effectively transfer mass to the companion. However, observations of the red-nova remnants V838\,Mon and V4332\,Sgr suggest that some material does fall back to the central objects decades after the merger \citep{infall,Tylenda2009,Tylenda2015}.

\subsection{Alternative origins of the molecular remnant}
Accretion-powered ejections are not the only viable possibility explaining the origin of the jet-like outflows with varying orientations. One of the scenarios worth considering is erratic mass loss from a magnetic merged star. Merger products are expected to develop strong magnetic fields \citep{SokerTylendaMagnetic,Antonini,Schneider} which enhance mass-loss rates from such a newly-formed star. In some merger scenarios, magnetic winds are considered an important sink of the excess angular momentum stored in the envelope of a merger product \citep[cf.][]{Braithwaite}. Mass loss from such an object is expected to be inhomogeneous and erratic in time and orientation. 
Such an expanding outflow may be deflected into bipolar lobes by a disk or a torus-like structure formed during the merger and placed in the orbital plane of the former binary. Observational and theoretical verification of such a scenario would however be very difficult.

\subsection{CK\,Vul analogs among PPNe?}
We have outlined many similarities between CK\,Vul's nebula and PPNe. Are there objects classified as post-AGB or PPNe, but that in fact of a similar nature as CK\,Vul? We think that is possible. One prominent distinction between CK\,Vul remnant and bone fide PPNe is the much lower bolometric luminosity of 12--60 L$_{\sun}$ compared to $10^4$\,L$_{\sun}$ typical for true post-AGB stars. Some PPNe and PNe, although called post-AGB, are actually thought to be post-RGB systems \citep[cf.][]{OOmen2019,Kamath} with luminosities close to the upper limit of CK\,Vul's luminosity. These objects may be a good starting point in searching for a CK\,Vul analog. We expect that post-merger systems would have systematically higher masses of the circumstellar material than true post-RGB stars with undisturbed evolution. 

For many genuine post-AGB PPNe, the distance and the luminosity are unknown or poorly constrained --- only a few have well known parallaxes. Objects that appear on images like PPNe are often assumed to have luminosities of $\approx$10$^4$\,L$_{\sun}$ even though their true evolutionary status is unknown  \citep[e.g., see discussion in][]{FrostyLeo}. There is a chance that relatives of CK\,Vul will be found among such mis-identified PPNe once their true luminosities are known better. Determining the distances for these heavily-embedded stars is difficult, as the visual extinction may be as high 100\,mag,  but is not impossible. 

Despite the similarities to PPNe CK\,Vul is unique owing to the well known eruptive history going back to the 17th century. Many PPNe have kinematical ages of a few hundred years, too, and are suspected to have undergone an explosive-type event \citep[e.g.,][]{Lee, Schmidt}. 
If the lobe ejections were associated with bright optical outbursts, as that of CK\,Vul and of other red novae, observational records from the last few centuries should have reported similar eruptive events (at least for a few younger PPNe in the northern hemisphere). No such records are known today. Perhaps the outburst were overlooked because they were not bright in the visual. Merger progenitors can enshroud the system in dust years prior to the merger \citep{Tylenda2011,Pejcha}. For certain geometries, this would make the outburst essentially unobservable at visual wavelengths but bright in the infrared. Dusty infrared transients have been recognized only very recently \citep{TylendaBLG,Jencson} and their link to PPNe is indeed intriguing. There is a chance that among objects classified today as PPNe are merger remnants resembling CK\,Vul.  

\section{Summary}
Through 3D modeling of the optical nebula of CK\,Vul, we revised the distance to the object to $>$2.6\,kpc which makes the stellar remnant and Nova 1670 significantly more luminous than assumed so far. At the revised distance, the hourglass structure has an enormous size, of >0.9\,pc, and its outermost parts move at very high speeds, >1400\,\kms. Using ALMA data from Paper\,I, we also reconstructed the 3D spatio-kinematical structure of the molecular remnant of CK\,Vul. Neutral species and ions fill walls of complex structures that could not have been created in a single mass ejection episode. Gas is traced in a confined ranges of the azimuthal angle. The distribution of molecular ions appears especially intricate, implying more complex ejections or an extra interaction mechanism that remains unidentified. Within the model uncertainties, the molecular remnant could have been created in the 17th century eruption or a century later. The molecular bipolar outflows carry out only $\approx$40\% of the total mass of the cool molecular gas.  With de-projected speeds reaching 470\,\kms, however, they contain most of the kinetic energy and linear momentum of the molecular remnant. These physical parameters of CK\,Vul's outflows are very similar to those of classical PPNe outflows. By analogy to current ideas of PPN formation, we propose that the elaborate molecular structures in CK\,Vul were created by episodic jets with a varying orientation and we speculate that there may be a binary system in the remnant center. We postulate that post-merger remnants similar to CK\,Vul could be hiding among sources incorrectly classified as post-RGB and post-AGB objects. The current study does not challenge the interpretation of CK\,Vul as a remnant of a stellar merger that involved an RGB star \citep{NatAstr}.

\begin{acknowledgements}
T.K. acknowledges funding from grant no 2018/30/E/ST9/00398 from the Polish National Science Center. W.S. acknowledges a support from UNAM DGAPA PASPA. R.T. acknowledges a support from grant 2017/27/B/ST9/01128 financed by the Polish National Science Center. M.H. thanks the Ministry of Science and Higher Education (MSHE) of the Republic of Poland for granting funds for the Polish contribution to the International LOFAR Telescope (MSHE decision no. DIR/WK/2016/2017/05-1) and for maintenance of the LOFAR PL-612 Baldy (MSHE decision no.~59/E-383/SPUB/SP/2019.1).
This paper makes use of the following ALMA data: ADS/JAO.ALMA\#2017.1.00999.S, ADS/JAO.ALMA\#2017.A.00030.S, ADS/JAO.ALMA\#2016.1.00448.S, and ADS/JAO.ALMA\#2015.A.00013.S. ALMA is a partnership of ESO (representing its member states), NSF (USA) and NINS (Japan), together with NRC (Canada), MOST and ASIAA (Taiwan), and KASI (Republic of Korea), in cooperation with the Republic of Chile. The Joint ALMA Observatory is operated by ESO, AUI/NRAO and NAOJ. The National Radio Astronomy Observatory is a facility of the National Science Foundation operated under cooperative agreement by Associated Universities, Inc.
The Submillimeter Array is a joint project between the Smithsonian Astrophysical Observatory and the Academia Sinica Institute of Astronomy and Astrophysics and is funded by the Smithsonian Institution and the Academia Sinica.
This work has made use of data from the European Space Agency (ESA) mission {\it Gaia} (\url{https://www.cosmos.esa.int/gaia}), processed by the {\it Gaia} Data Processing and Analysis Consortium (DPAC, \url{https://www.cosmos.esa.int/web/gaia/dpac/consortium}). Funding for the DPAC
has been provided by national institutions, in particular the institutions participating in the {\it Gaia} Multilateral Agreement.
Based on observations obtained at the international Gemini Observatory, a program of NSF’s NOIRLab, which is managed by the Association of Universities for Research in Astronomy (AURA) under a cooperative agreement with the National Science Foundation. on behalf of the Gemini Observatory partnership: the National Science Foundation (United States), National Research Council (Canada), Agencia Nacional de Investigaci\'{o}n y Desarrollo (Chile), Ministerio de Ciencia, Tecnolog\'{i}a e Innovaci\'{o}n (Argentina), Minist\'{e}rio da Ci\^{e}ncia, Tecnologia, Inova\c{c}\~{o}es e Comunica\c{c}\~{o}es (Brazil), and Korea Astronomy and Space Science Institute (Republic of Korea). Several figures in this paper were prepared with {\tt matplolib} \citep{matplotlib} and with SAO DS9. 
\end{acknowledgements}

\end{document}